\documentclass[a4paper,11pt]{article}
\pdfoutput=1 % if your are submitting a pdflatex (i.e. if you have
\usepackage{jheppub} % for details on the use of the package, please
\usepackage[T1]{fontenc} % if needed
\usepackage[utf8]{inputenc}
\usepackage{amsfonts}
\usepackage{amsmath,url,hyperref}

\usepackage{caption}
\usepackage{subcaption}
\usepackage[dvipsnames]{xcolor}

\title{Picard-Lefschetz decomposition and Cheshire Cat resurgence in $3$D \texorpdfstring{$\mathcal{N}=2$}{TEXT} field theories.}
\preprint{DCPT-19/27}

\author{Daniele Dorigoni}
\author{and Philip Glass}
\affiliation{Department of Mathematical Sciences, Durham University,\\Lower Mountjoy, Stockton Road, Durham, DH1 3LE, UK}

\emailAdd{daniele.dorigoni@durham.ac.uk}
\emailAdd{philip.glass@durham.ac.uk}

\abstract{We study three dimensional \texorpdfstring{$\mathcal{N}=2$}{TEXT} supersymmetric abelian gauge theories with various matter contents living on a squashed sphere. In particular we focus on two problems: firstly we perform a Picard-Lefschetz decomposition of the localised path integral but, due to the absence of a topological theta angle in three dimensions, we find that steepest descent cycles do not permit us to distinguish between contributions to the path-integral coming from (would-be) different topological sectors, for example a vortex from a vortex/anti-vortex. The second problem we analyse is the truncation of all perturbative expansions. Although the partition function can be written as a transseries expansion of perturbative plus non-perturbative terms, due to the supersymmetric nature of the observable studied we have that each perturbative expansion around trivial and non-trivial saddles truncates suggesting that normal resurgence analysis cannot be directly applied. 
The first problem is solved by complexifying the squashing parameter, which can be thought of as introducing a chemical potential for the global $U(1)$ rotation symmetry, or equivalently an omega deformation. This effectively introduces a hidden ``topological angle'' into the theory and the path integral can be now decomposed into a sum over different topological sectors via Picard-Lefschetz theory. The second problem is solved by deforming the matter content making manifest the Cheshire Cat resurgence structure of the supersymmetric theory, allowing us to reconstruct non-perturbative information from perturbative data even when these do truncate.}

\begin{document} 
\maketitle
\flushbottom

\section{Introduction}

In recent years Ecalle's resurgence theory \cite{Ecalle:1981} has been applied to an ever growing set of problems where the common denominator is the asymptotic nature of the perturbative expansion and how we can exploit its lack of convergence to reconstruct non-perturbative results out of perturbative data. For some recent introductions see \cite{Dorigoni:2014hea,Aniceto:2018bis} and references therein.

A particularly fortuitous class of examples where we can try to apply resurgence theory are supersymmetrically localisable field theories. 
Starting with Pestun's seminal work \cite{Pestun:2007rz} for $\mathcal{N}=4$ and $\mathcal{N}=2$ theories on $S^4$, many quantities like partition functions and Wilson loops have been computed exactly using supersymmetric localisation; see \cite{Cremonesi:2014dva} for a pedagogical introduction and a more complete set of references.

This method is very general and can be applied to theories living on different manifolds, in various numbers of dimensions, and with various amounts of supercharges. For example one can consider $\mathcal{N}=2$ theories on a squashed $S^4$ \cite{Hama:2012bg}, or in three dimensional $\mathcal{N}=2$ on a round \cite{Hama:2010av} or squashed sphere \cite{Hama:2011ea}, or similarly going to two dimensional $\mathcal{N}=(2,2)$ theories on a sphere \cite{Benini:2012ui,Doroud:2012xw} or an ellipsoid \cite{Gomis:2012wy}.

Importantly in all these cases the exact localised partition functions and other observables can be written as a perturbative part plus non-perturbative sectors, i.e. what is usually referred to as a \textit{transseries} \cite{Edgar:2008ga}. 
Hence a very natural question is whether or not one can apply resurgent methods to these quantities and reconstruct the complete answers from the purely perturbative data.
This question was analysed in \cite{Russo:2012kj,Aniceto:2014hoa,Honda:2016mvg,Honda:2016vmv,Fujimori:2018nvz} and it was realised that in supersymmetric field theories the resurgence story is not as straightforward.

Although in some examples one can use the perturbative data to reconstruct the presence of new, complexified saddle points to the path-integral \cite{Honda:2017qdb}, in many others these authors realised that perturbation theory seem to be oblivious to non-perturbative sectors which we know must be present from the exact localised formulae, or even worse (or better depending on the point of view) cases in which all the perturbative expansions have actually finite radius of convergence and for which the resurgence programme seems completely doomed to fail.

One does not need to go to complicated theories to construct an example of this: just in standard supersymmetric quantum mechanics \cite{Witten:1981nf} we know that the ground state energy will vanish in perturbation theory because of bosonic/fermionic cancellations; however non-perturbative effects can lift the vacuum energy.
In \cite{Dunne:2016jsr,Kozcaz:2016wvy} the authors studied precisely these supersymmetric quantum mechanical models and proved that this ``lack'' of resurgence is an extremely fine tuned phenomenon which they called Cheshire Cat resurgence.
Just like the eponymous cat, the full resurgence body is still there; one just need to introduce a small deformation to the theory to immediately obtain an asymptotic, factorially growing perturbative expansion.
In this deformed theory we can use the full resurgence machinery to extract the non-perturbative sectors out of perturbative data and once we send the deformation to zero the perturbative series will truncate (or become convergent) while the non-perturbative terms will still be there, thus the grin will linger on.

Following \cite{Dunne:2016jsr,Kozcaz:2016wvy}, in \cite{Dorigoni:2017smz} we applied a similar deformation to the $\mathcal{N}=(2,2)$ supersymmetric $\mathbb{CP}^{N-1}$ on $S^2$.
The reason why perturbation theory seems to be oblivious to the non-perturbative sectors is precisely the same and a small deformation restores immediately the full body of the Cheshire Cat resurgence. The supersymmetrically localised partition function is sitting at a very special point in theory space thus hiding the full resurgence structure. This is the reason why in some cases of \cite{Russo:2012kj,Aniceto:2014hoa,Honda:2016mvg,Honda:2016vmv} there seemed not to be any resurgence at all.

When considering path-integrals, complementary to resurgence analysis \cite{Dunne:2015eaa} is the Picard-Lefschetz theory or complexified Morse homological decomposition in steepest descent contours \cite{Pham1983}, see also \cite{Witten:2010cx,Witten:2010zr,Harlow:2011ny}. 
The key idea is that one has to deform the path-integral contour of integration into a suitable complexification of field space. Associated to each complex saddle point there is a privileged, steepest descent contour of integration, usually called a Lefschetz thimble, and at a generic value of the coupling constant one can rewrite the original contour as a linear combination of these thimbles with integer coefficients\footnote{Note that although when dealing with finite dimensional integrals the intersection numbers will always be integers, in infinite dimensions this is not necessarily guaranteed, see for example \cite{Gukov:2016njj}.}, i.e. intersection numbers.

\begin{figure}[ht]
\begin{center}
\includegraphics[width=0.8\textwidth]{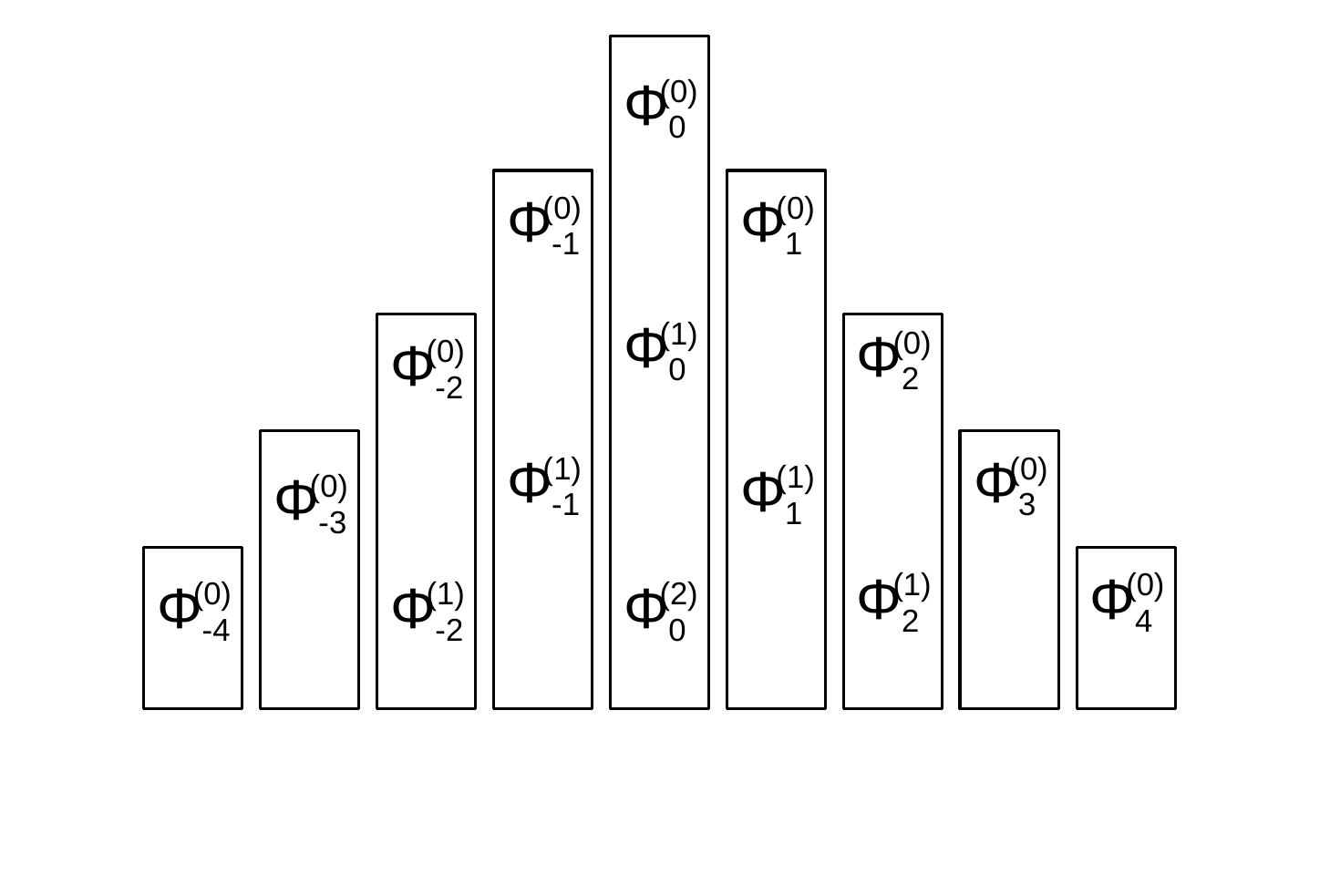} 
\caption{The Resurgence triangle. The $k^{th}$ non-perturbative contribution part of the $N^{th}$ topological sector is denoted schematically by $\Phi^{(k)}_N$. Resurgence theory allows us to reconstruct from any $\Phi^{(k)}_N$ all the other contributions in the \textit{same} column, i.e. $\Phi^{(k')}_N$.}
\label{Resurgencetriangle}
\end{center}
\end{figure}

The link between resurgence and Picard-Lefschetz decompositions comes from Stokes phenomenon. For special arguments of the complexified coupling constant, i.e. Stokes directions, we have that a thimble can connect different saddles. This is usually forbidden given the fact that the imaginary part of the action is constant along a thimble, i.e. they are stationary phase contours. Across a Stokes direction some of the thimbles will undergo non-trivial monodromies and the aforementioned intersection numbers will jump. Simultaneously resurgence analysis tell us that the resummation of the asymptotic series around the saddles involved will also jump and these two discontinuities, of intersection numbers and resummations, are tightly related.

Whenever the theory in question contains a topological $\theta$ angle this will contribute to the classical action of the various saddle points by an imaginary part weighted by $\theta$ times the topological number. Thus even before complexifying the coupling constant (note that one should not confuse the imaginary part of the complexified coupling constant with the theta angle) we have that steepest descent paths can only connect saddles coming from the same topological sector. Thus generally whenever a theta angle is present the path integral will first split into a sum over topological sectors; then upon complexification of the coupling constant, resurgence theory will allow us to relate the perturbative series in a given topological sector to non-perturbative saddles in the same topological sector, e.g. from the purely perturbative expansion around the trivial vacuum we are able to reconstruct instanton-anti-instanton events. 
The resurgence structure of the theory arranges itself in what is called the resurgence triangle \cite{Dunne:2012ae} as in Figure \ref{Resurgencetriangle}. From an asymptotic series around a saddle point (perturbative and non-) in a given topological sector we can calculate all the other saddles in the same sector.

For the present work we are interested in $3$-d $\mathcal{N}=2$ gauge theories. As mentioned above this class of theories is amenable to localisation on $S^3$ \cite{Hama:2010av} and,
when a Chern-Simons term is present, their partition functions can be directly written \cite{Honda:2016vmv} in the form of a resurgent transseries in terms of a small coupling $g=1/k\ll1$ given by the inverse of the Chern-Simons level $k$. 

In \cite{Fujimori:2018nvz}\footnote{We thank Tatsuhiro Misumi for making us aware of this interesting and relevant work.} the authors performed the complete resurgent analysis and thimble decomposition for $3$-d $\mathcal{N}=2$ Chern-Simons matter theories showing that, as one varies the argument of the coupling $g=1/k$, the thimble decomposition of the path-integral exhibits Stokes phenomenon. As expected the ambiguities in resummation of the Borel transform are directly related to the jump in thimbles attached to non-perturbative saddles. 
Furthermore the analysis of these authors provided a nice interpretation of these non-perturbative effects as contributions coming from new supersymmetric solutions \cite{Honda:2017qdb} living in a complexification of field space but not on the original path-integral contour.

In this work we continue the studies of \cite{Fujimori:2018nvz} by considering abelian gauge theories without a Chern-Simons term for which both asymptoticity and topological angle will turn out to be absent. 
At first it looks like the resurgence structure found in \cite{Honda:2016vmv,Fujimori:2018nvz} disappears completely, and although we have these complex non-perturbative saddles \cite{Honda:2017qdb}, their classical actions will nonetheless be real, hence the Picard-Lefschetz decomposition of \cite{Fujimori:2018nvz} somehow becomes degenerate. The same holds even when considering a squashed sphere \cite{Hama:2011ea} with squashing parameter $b>0$. However, by complexifying the squashing parameter $b=e^{i\theta}$ we will be able to identify ``would-be'' different topological sectors, i.e. we can distinguish from the thimble point of view the topologically trivial sector from a vortex and an anti-vortex.

The complexification of the squashing parameter can be seen as the introduction of a chemical potential for the $U(1)$ rotation of the $S^2$ where the vortices are living on when we write the $S^3$ as a Hopf fibration. It is also interesting to notice that since the building blocks to compute the $3$-d $\mathcal{N}=2$ partition functions are directly related \cite{Hosomichi:2010vh,Hama:2011ea} to the structure constants in $2$-d Liouville with central charge $c=1+6(b+b^{-1})^2$   we have that the complexification $b=e^{i\theta}$ interpolates precisely between space-like and time-like Liouville.

However, regardless of the interpretation, the important point is that complexifying the squashing parameter generates a topological angle which was hidden before, and this is very reminiscent of the hidden topological angle studied in \cite{Behtash:2015kna,Behtash:2015kva,Behtash:2015zha,Behtash:2015loa,Dunne:2016jsr}.
We can thus introduce a Cheshire Cat deformation very similar to \cite{Dorigoni:2017smz} and restore the asymptotic nature of the perturbative series around each saddle. This allows us to use the full resurgence machinery to reconstruct from just one element in a column of the resurgence triangle all other elements in the same column, i.e. in the same topological sector.

This topological decomposition combined with the known \cite{Pasquetti:2011fj} vortex/anti-vortex factorisation of the partition function introduces an extra structure on top of resurgence and we can now calculate data from one column of the resurgence triangle and relate it to different columns. An example of this ``horizontal'' move on the resurgence triangle is given by the Dune-\"Unsal relation in quantum mechanics (see \cite{Dunne:2013ada,Dunne:2014bca,Gahramanov:2015yxk}). 

The paper is organised as follows. Firstly in Section \ref{TheTheory} we will briefly give an overview of $\mathcal{N}=2$ supersymmetric field theories on a squashed $3$-sphere and present the localised partition functions for different matter contents.

In Section \ref{PLsection} we will perform a Picard-Lefschetz decomposition of the localised path-integral and show how a complexification of the squashing parameter $b$ will give rise to a hidden topological angle, allowing us to decompose the theory in a resurgence triangle structure. We will also discuss the physical interpretation of this complexificiation and the arising of Stokes phenomenon in the Picard-Lefschetz decomposition. 

We continue in Section \ref{resurgenceanalysis} with the analysis of the localised path integral using Cheshire Cat resurgence methods. Similarly to the $2$-d case we will see that for the original theory there is no asymptotic, factorially growing perturbative series. However after introducing a suitable deformation the asymptotic nature of perturbation theory is reinstated and the resurgence framework can finally be applied. We also discuss how to connect the non-perturbative data to the perturbative data, and at the very end smoothly continue all of our results back to the undeformed original theory while retaining the non-perturbative information acquired from perturbation theory.

We will remark on additional structures going beyond resurgent theory in Section \ref{dunneunsalsection}. These additional structures are very common in supersymmetrically localised partition functions in various dimensions, and they are very reminiscent of the quantum mechanical Dunne-\"Unsal relations \cite{Dunne:2014bca,Dunne:2013ada,Gahramanov:2015yxk}, which allow us to move ``horizontally'' in the resurgence triangle, thus deriving  data in different topological sectors just by analysing the perturbative one.
Finally we draw some conclusions and remarks on future works in Section \ref{conclusion} and present some useful identities for the double sine function in Appendix \ref{doublesine}.

\section{\texorpdfstring{$\mathcal{N}=2$}{TEXT} theories on squashed \texorpdfstring{$S^3$}{TEXT}}\label{TheTheory}

It will be useful to recall some facts about $\mathcal{N}=2$ gauge theories on $S^3$ and how to calculate their partition functions from localisation methods. For a nice review of 3D $\mathcal{N}=2$ theories see \cite{Aharony:1997bx}, and for all the details of the localisation calculation see \cite{Hama:2011ea} and \cite{Fujitsuka:2013fga}.

Three dimensional field theories with $\mathcal{N}=2$ supersymmetry have 4 real supercharges, and can be seen as a reduction of $4$-d $\mathcal{N}=1$ theories down to three dimensions. Vector multiplets contain a vector field, two Weyl-fermions and a scalar, while chiral or anti-chiral matter multiplets are made of two complex scalars, and two Weyl-fermions.

We will be interested in theories with an abelian vector multiplet and various chiral and/or anti-chiral matter multiplets charged under the gauge symmetry; hence supersymmetric Lagrangians will then contain a Yang-Mills part, a matter part, a Fayet-Iliopoulos term (\textit{FI} in what follows) with parameter $\xi$, and finally a Chern-Simons term (which we will not be concerned with). 
In the rest of this paper we will express everything in terms of the FI parameter, and thus the weak coupling expansion will correspond to large $\xi\gg 1$, and the strong coupling expansion will have $\xi\sim 0$.

We will be concerned with theories defined on a squashed $3$-sphere which can be embedded in $4$-d via
\begin{eqnarray}
\frac{b^2}{r^2}\left(x_0^2+x_1^2\right)+\frac{1}{b^2 r^2}\left(x_2^2+x_3^2\right)=1\; ,
\end{eqnarray}
and will be denoted as $S_b^3$ where $b$ is our squashing parameter, and $r$ is the radius of the sphere that we will set to $1$ in appropriate units.
Note that in here $b$ is thought of as a positive real number and $b=1$ corresponds to the round sphere case.

The partition function can be computed following the procedure laid out in \cite{Pestun:2007rz}, and performed in \cite{Hama:2011ea} and \cite{Fujitsuka:2013fga}. In the case where we do not turn on any mass parameters, the squashed $S^3$ partition function can be written as
\begin{eqnarray}
Z_{S_b^3}=\int d\hat{x} \,e^{2\pi i \xi\,\mbox{Tr}(\hat{x})}Z_{vec}(\hat{x})Z_{matter}(\hat{x}) \; ,
\end{eqnarray}
where the integral is over the Cartan subalgebra of the gauge group.
The parameter $\xi$ is the usual FI term, and the one-loop determinants are for the vector multiplet
\begin{eqnarray}
Z_{vec}(\hat{x})=\prod\limits_{\alpha\in \Delta_+}\sinh(\pi b \alpha(\hat{x}))\sinh(\pi b^{-1} \alpha(\hat{x})) \;\;\; ,
\end{eqnarray}
while for the chiral/anti-chiral multiplets
\begin{eqnarray}
Z_{matter}(\hat{x})=\prod\limits_{w\in R}s_b\left(\frac{iQ}{2}(1-\Delta)-w(\hat{x})\right) \;\;\; .
\end{eqnarray}
Note here we have used $\Delta_+$ to denote the positive roots, $w$ to denote the weights in representation $R$ of the matter multiplet, $b$ is once again the squashing parameter and $Q=b+1/b$, and $\Delta$ is the R-charge of the scalar in the chiral multiplet. For abelian gauge theories the vector multiplet one-loop determinant will simply be one. 

The matter one-loop determinants can be all written in terms of the double sine function $s_b(x)$ presented here in terms of an infinite product
\begin{eqnarray}\label{eq:sbproduct}
s_b\left(x\right) &=& \prod_{m,n\geq0}\frac{(mb+n/b+Q/2-ix)}{(mb+n/b+Q/2+ix)}\;\;\; .
\end{eqnarray}
This function is closely related to the hyperbolic gamma function \cite{Rujisenaars,DeBult} and the multiple sine function \cite{Kurokawa}, and we present some of its properties in Appendix \ref{doublesine}. 
The only property of $s_b(x)$ we want to stress here is that for generic $b$ this function has simple zeroes on the lattice $\Lambda_+ =-iQ/2 -i b \,\mathbb{Z}_{\geq 0} -i/b\, \mathbb{Z}_{\geq 0}$ and simple poles on the lattice  $\Lambda_- = +i Q/2 +i b\, \mathbb{Z}_{\geq 0} +i/b\, \mathbb{Z}_{\geq 0}$.
In the physical squashing limit $ b\in \mathbb{R}^+$ the poles and zeroes of the one-loop determinant fall on the imaginary axis and correspond to the appearance of bosonic and fermionic, respectively, massless states.

For the sake of simplicity in the present work we will only be concerned with $U(1)$ gauge theories, though all our conclusions should carry over to the non-abelian case quite simply. For us therefore the vector multiplet one-loop determinant $Z_{vec}$ will always be equal to $1$, and the partition function will depend on the FI parameter $\xi$, the squashing $b$, and the number of chiral, $N_c$, and anti-chiral, $N_a$, multiplets\footnote{In the present paper we will also work with $N_c-N_a$ an even number as to avoid having to introduce a bare Chern-Simons term to cancel the parity anomaly \cite{Redlich:1983kn,Redlich:1983dv}.}. For this class of theories we have
\begin{eqnarray}\label{localisedpartition}
Z_{S^b_3}^{(N_c,N_a)}(\xi)=\int_\Gamma dx\, e^{2\pi i \xi x}\,\frac{\prod\limits_{i=1}^{N_c}s_b(x+iQ/2)}{\prod\limits_{i=1}^{N_a}s_b(x-iQ/2)}\;\;\;,
\end{eqnarray}
where the contour $\Gamma$ runs along the real $x$ axis and circles around the origin passing in the lower complex $x$ half-plane. Note that we chose the R-charge of the scalars to be $\Delta=0$.

These integrals can be calculated by closing the contour in the upper half plane and picking up contributions from all the poles thus leaving a sum over the residues of these poles. General results with non-zero vector and axial masses can be found in \cite{Pasquetti:2011fj}. 

We will shortly analyse the Picard-Lefschetz decomposition and the Cheshire Cat deformation of these types of theories, and to this end it will be useful to better understand the analytic properties of the integrands and their dependence on the matter content and squashing parameter. For this reason we will now discuss some particular examples in more detail.

\subsection{Round sphere}

We start off by considering the theory on the round $S^3$, i.e. $b=1$, with any matter content. Using equations (\ref{eq:sbproduct}) and (\ref{localisedpartition}) we have that the partition function on the sphere is given by
\begin{eqnarray}
Z_{S^3}^{(N_c,N_a)}(\xi)=\int_\Gamma dx\; e^{2\pi i \xi x}\prod\limits_{m,n\geq 0}^{\infty} \left(\frac{m+n+2-ix}{m+n+ix}\right)^{N_c}\left(\frac{m+n+2+ix}{m+n-ix}\right)^{N_a}.
\end{eqnarray}
We can now rearrange the product by defining $L=m+n$, and realising that for fixed $L\in\mathbb{N}$ we have $L+1$ distinct pairs $(m,n)$ such that $L=m+n$, so we can write
\begin{eqnarray}
Z_{S^3}^{(N_c,N_a)}(\xi)&=&\int_\Gamma dx\; e^{2\pi i \xi x}\prod\limits_{L=0}^{\infty} \left(\frac{L+2-ix}{L+ix}\right)^{N_c(L+1)}\left(\frac{L+2+ix}{L-ix}\right)^{N_a(L+1)} .\;\;\;\;\;\;
\end{eqnarray}

One can evaluate this infinite products using zeta-regularisation (see for example the Appendix of \cite{Dorigoni:2017smz}) or alternatively for the non-chiral theory $N_c=N_a=N$ one can use equation (\ref{eq:functional2}) to obtain
\begin{eqnarray}
Z_{S^3}^{(N,N)}(\xi)&\notag=&\int_\Gamma dx \; e^{2\pi i \xi x}\left(\frac{1}{2\sinh(\pi x)}\right)^{2N} = \frac{(-1)^N}{\Gamma(2N)} \sum_{n=0}^\infty  e^{-2\pi n \xi}\,\xi
\prod_{k=1}^{N-1}(\xi^2+k^2)\\
&\label{s3partitionfunction}=& \frac{(-1)^N}{\Gamma(2N)}\frac{\xi}{1-e^{-2\pi \xi}} \prod_{k=1}^{N-1}(\xi^2+k^2)\,,
\end{eqnarray}
which can be obtained as the limit $b\to 1 $ and vanishing vector and axial masses of the general expression obtained in \cite{Pasquetti:2011fj}. 

Note that the partition function has simple poles at $\xi =  i k$ for $k\in \mathbb{Z}$; however for $k \in \{\pm 1,...,\pm (N-1)\}$ these are cancelled by the simple zeroes coming from the product. This can be understood from the mirror theory \cite{Aharony:1997bx,deBoer:1997ka} as due to the presence of a single bosonic zero mode for the monopole operators for $\xi =  i k$ with $k\in \mathbb{Z}$. However when $k \in \{\pm 1,...,\pm (N-1)\}$ the monopole operators acquire also a fermionic zero mode thus giving a finite, non-zero, contribution\footnote{We thank Stefano Cremonesi for clarifications on this point.}.

A particular case that will shortly be useful is when $N_c=N_a=1$ and the above expression simplifies to
\begin{eqnarray}\label{s3partitionfunction1}
Z_{S^3}^{(1,1)}(\xi)&=&\int_\Gamma dx \, e^{2\pi i \xi x}\,\frac{1}{4\sinh(\pi x)^2}= -\frac{\xi}{1-e^{-2\pi \xi}}\,.
\end{eqnarray}

It is manifest both in the above equation as well as in the general case (\ref{s3partitionfunction}) that the $S^3$ partition function takes the form of a transseries for which the perturbative expansion in each non-perturbative sector truncates because of supersymmetry after $2 N -1$ orders, where again $N = N_c=N_a$.

\subsection{Non-Chiral theory on squashed \texorpdfstring{$S^3$}{TEXT}}

In the case of the non-chiral theory, i.e. when $N_c=N_a$, on the squashed $3$-sphere we have the identity given in equation (\ref{eq:functional2}), which enables us to write
\begin{eqnarray}
\frac{s_b\left(x+iQ/2\right)}{s_b\left(x-iQ/2\right)}=\frac{1}{4\sinh(\pi xb)\sinh(\pi x/b)} \;\;\; .
\end{eqnarray}
Hence we can write the partition function for the non-chiral theory as
\begin{eqnarray}\label{eq:SquashedChiral}
Z^{(N,N)}_{S^3_b}(\xi)=\int_\Gamma dx\; e^{2\pi i x \xi}\left(\frac{1}{4\sinh(\pi xb)\sinh(\pi x/b)}\right)^N\;\;\; .
\end{eqnarray}
For $N=1$ it is fairly simple to compute the residues and obtain
\begin{equation}\label{eq:SquashedChiral1}
Z^{(1,1)}_{S_3^b}(\xi)=-\xi+\frac{1}{2}\sum_{n=1}^\infty (-1)^n \left[ e^{-2\pi n \xi b} \csc(n \pi b^2) b+e^{-2\pi n \xi/ b}\frac{\csc(n \pi/ b^{2})}{b}\right]\,,
\end{equation}
which reproduces (\ref{s3partitionfunction1}) when we take the $b\to 1$ limit.
As is well known the reason for these two different types of exponentially suppressed corrections comes from the fact that vortices are finite action solutions in $2$-d and finite energy solutions in $3$-d. However since our $3$-d manifold can be seen as an $S^1$ fibration over $S^2$ we can understand the $3$-d vortex action as its energy timed by the length of the $S^1$ fibre, hence precisely either $ 2\pi \xi n \times b$ or $ 2\pi \xi n \times b^{-1}$ depending on which $S^1$ we are fibering.
We note also that due to the supersymmetric nature of the observable under consideration the perturbative expansion in $\xi \gg 1$ around the vacuum, as well as all the non-perturbative sectors, does truncate after finitely many orders.

\subsection{Chiral theory on squashed \texorpdfstring{$S^3$}{TEXT}}

For $N_c \neq N_a$ on the squashed $3$-sphere things are a bit harder and one has to introduce the q-Pochhammer symbol, denoted by $(a;q)_\infty$, to obtain a regularised formula for $s_b(x)$ given in equation (\ref{qpochsb}). From this we can write the partition function as
\begin{eqnarray}
Z^{(N_c,N_a)}_{S^3_b}(\xi)=\int_\Gamma &&dx\; e^{2\pi i x \xi} \left(e^{-i\pi\frac{(x+iQ/2)^2}{2}}\frac{\left(e^{2\pi bx+2 \pi ib^2};e^{2\pi ib^2}\right)_\infty}{\left(e^{2\pi x/b};e^{-2\pi i/b^2}\right)_\infty}\right)^{N_c}  \\
&&\nonumber \left(e^{-i\pi\frac{(x-i Q/2)^2}{2}}\frac{\left(e^{2\pi b x};e^{2\pi ib^2}\right)_\infty}{\left(e^{2\pi x/b-2\pi i/b^2};e^{-2\pi i/b^2}\right)_\infty}\right)^{-N_a} \alpha^{N_c-N_a}\; ,
\end{eqnarray}
where again $Q= b+1/b$ and we introduce the constant $\alpha = \exp( -i \pi\frac{Q^2-2}{24})$.
Note that the q-Pochhammer $(a;q)_\infty$ has a natural boundary of analyticity at $\vert q\vert =1$. We will shortly see that our complexification of the squashing parameter $b\to e^{i \theta}$ will bring us to work within the unit disk for q-Pochhammers.
This expression will be useful when analysing the Picard-Lefschetz decompositions.

\section{Picard-Lefschetz decomposition and hidden topological angle}\label{PLsection}

We start our analysis of the Picard-Lefschetz decomposition of the localised path-integral by considering first theories with a real squashing parameter $b>0$, and subsequently complexifying it.
As it will become clear later on the combination $\Theta= -i(b-1/b)$ will play the role of hidden topological angle, and hence, by abuse of notation, in this Section we will say that a give saddle belong to the $N^{th}$ topological sector if the imaginary part of its action goes like $N \Theta$; for example the perturbative saddle and the vortex-anti-vortex saddle both have $N=0$ while the vortex and the anti-vortex have $N=1$ and $N=-1$ respectively.

For concreteness let us consider a theory with 1 chiral and 1 anti-chiral multiplet on a round $S^3$, i.e. $b=1$. The partition function is given by equation (\ref{s3partitionfunction1}) and it is simple to note that the integrand has double order poles at $x=in$ for $n\in\mathbb{Z}$. Let us now look at this path integral from a Picard-Lefschetz point of view. To this end we exponentiate the one-loop determinant and write the integrand in terms of an effective action
\begin{eqnarray}
Z_{S^3}^{(1,1)}(\xi)&=&\int_\Gamma dx \;e^{2\pi i x \xi}\frac{1}{4\sinh(\pi x)^2} = \int_\Gamma dx \;e^{-S_{eff}^{(1,1)}(x)} \;\;\; , \nonumber \\
S_{eff}^{(1,1)}(x) &=& -2\pi i\xi x +2\log(2\sinh(\pi x))\;\;\; .\label{eq:Seff11}
\end{eqnarray}
The idea behind Picard-Lefschetz decomposition (we refer the reader to the nice expositions in \cite{Witten:2010cx,Witten:2010zr} for a more detailed account) is to use the effective action $S_{eff}$, or rather its real part, as a Morse function to construct a set of privileged contours living in the complexified field space $x\in\mathbb{C}$, called Lefschetz thimbles or alternatively steepest descent contours, with some crucial properties:
\begin{itemize}
\item the imaginary part of the action is constant along the thimble (stationary phase);
\item at a generic point in parameter space there is a thimble attached to one and only one critical point of the effective action;
\item the real part of the action is monotonically increasing as we move away from the critical point along its associated thimble;
\item the original contour of integration $\Gamma$ can be decomposed as a linear combination with integer coefficients (intersection numbers) of thimbles.
\end{itemize}

These thimbles can be constructed as the solution to the Morse flow equation
\begin{align}
&\frac{d x(t)}{dt} \notag= \pm\overline{ \frac{\partial S_{eff}(x(t))}{\partial x(t)}}\,,\\
&\lim\limits_{t\to-\infty} x(t) = x_{cr}\qquad\qquad\mbox{with}\,\,\left.\frac{\partial S_{eff}(x)}{\partial x}\right\vert_{x=x_{cr}}=0\,.
\end{align}
The solution with the plus sign is usually called the J thimble associated to the critical point $x_{cr}$, or J cycle (also called unstable or downward manifold) for which as we just stressed we have that the real part of the effective action is monotonically increasing. The solution with the minus sign defines the dual thimble, which we will call the K thimble associated to the critical point $x_{cr}$ (stable or upward manifold) and along which the real part of the effective action is monotonically decreasing.
At a generic point in parameter space we have that the intersection number of a J cycle and a K cycle is non-zero if and only if they are both associated with the same critical point.
This will allow us to decompose the original contour of integration $\Gamma$ as $\Gamma= \sum_{\sigma} n_{\sigma} J_{\sigma}$ where the sum runs over all the complex critical points of $S_{eff}$ and the coefficient $n_{\sigma} = (K_{\sigma} ,\Gamma)$ is just the intersection number of the contour $\Gamma$ with the K thimble attached to the critical point $\sigma$. 

\begin{figure}[ht]
 
\begin{subfigure}{0.5\textwidth}
\includegraphics[width=0.9\linewidth, height=6cm]{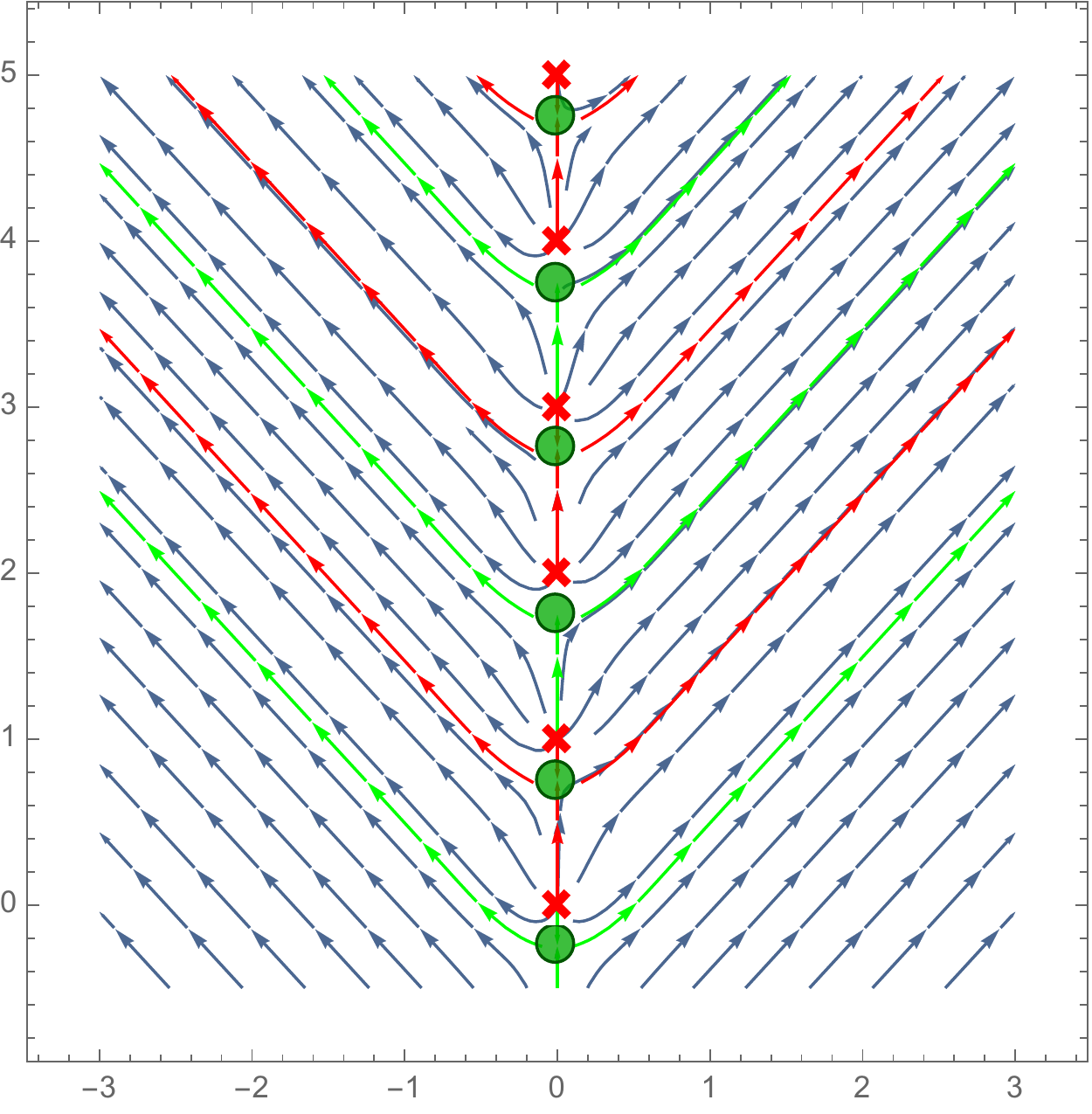} 
\caption{The flow in the upper half plane.}
\label{fig:subim1}
\end{subfigure}
\begin{subfigure}{0.5\textwidth}
\includegraphics[width=0.9\linewidth, height=6cm]{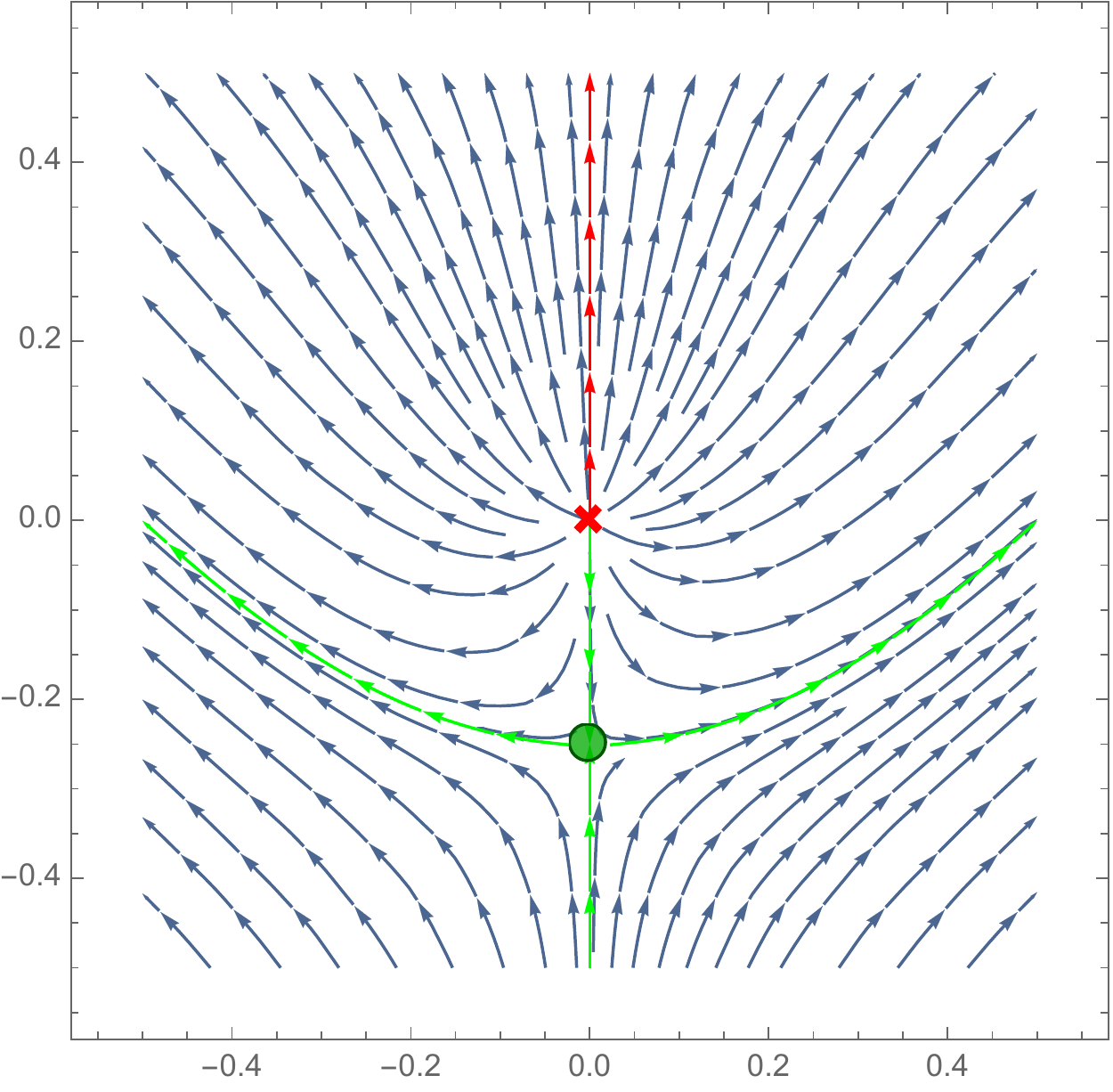}
\caption{Perturbative saddle and its nearest pole.}
\label{fig:subim2}
\end{subfigure}
 
\caption{Morse flow for the theory with one chiral and one anti-chiral, $b=1$, $\xi=1$. The green circles are saddle points while the red crosses are poles of the effective action. From each saddle the downward flows (the J cycles) go off to $\infty$ while the upward flows (K cycles) flow vertically until they hit a pole.}
\label{spheredecomposition}
\end{figure}

Following the idea outlined above we now try and perform the Picard-Lefschetz decomposition of the integration contour $\Gamma$ using the Morse flow induced by $S_{eff}^{(1,1)}(x)$ for the example above (\ref{eq:Seff11}). Since the effective action is basically the logarithm of the one-loop determinant we have that both zeroes and poles of the one-loop determinant will produce singularities of the effective action. 
Since we are interested in the $\xi \gg 1$ expansion of the path integral we have that each one of the saddle points will live close to each one of the singularities of the effective action (i.e. zeroes and poles of the one-loop determinant) and steepest descent and ascent cycles can now terminate at singular points of the effective action.
This is shown in Figure \ref{spheredecomposition}.

A very similar analysis was already carried out for $3$-d $\mathcal{N}=2$ Chern-Simons matter theories in \cite{Fujimori:2018nvz}, see in particular equation (II.21) and our (\ref{eq:Seff11}). Notice however some key differences with our results. In particular that when a Chern-Simons term is present the J thimble attached to the perturbative vacuum, noted with $\mathcal{J}_{\mbox{pt}}$ in
\cite{Fujimori:2018nvz}, passes through the lattice of saddles and poles, see for example their Figure 3, while in our Figure \ref{spheredecomposition}-(a) the perturbative thimble envelopes all the singularities and saddles.

In the extremely thorough analysis of \cite{Fujimori:2018nvz} the authors noted that as the real mass parameter is increased, or equivalently the argument of the coupling $g=1/k$ is varied, more and more non-perturbative thimbles cross the perturbative one and Stokes phenomenon take place, presented in their Figure 7 and 9. These jumps are directly correlated with the jumps in the resummation of the asymptotic expansion for small $g \ll1$.

At a first glance in our case none of these phenomena happen, the key difference being the absence of a Chern-Simons term. As we have already shown in Section \ref{TheTheory} the original contour of integration, which can be straightforwardly deformed to the perturbative thimble of Figure \ref{spheredecomposition}-(a), simply reduces the integral to a sum over residues hence not giving rise to any asymptotic perturbative expansion.

The presence of a Chern-Simons term changes completely the asymptotic form of the effective action for large Coulomb branch parameter $|x| \gg 1$ from the case at hand where $S_{eff}(x) \sim -2\pi i\xi_{\tiny{\mbox{eff}}} x $ to $S_{eff}(x) \sim -i x^2/ (4pi g)$ with $g=1/k$ being the inverse Chern-Simons level. As the level goes to zero, i.e. $g\to\infty$, we have a discontinuous jump in the asymptotic regions $\mbox{Re} \,S_{eff}(x)>0$, usually referred to as good regions \cite{Witten:2010cx}. As explained in details in \cite{Witten:2010cx} the J thimbles are non-compact and their tails must lie in the good regions. It is then the asymptotic behaviour of $\mbox{Re} \,S_{eff}(x)$ for $|x| \to \infty$ that dictates the topology of the thimbles. When a Chern-Simons level is present the good regions asymptote two quadrants $\mbox{Re} \,[ -i x^2/ (4pi g)] >0$ in the complex $x$-plane while in our case they asymptote a half plane $\mbox{Re} \,[2\pi i
\xi_{\tiny{\mbox{eff}}} x ]>0$. This is the reason why the perturbative thimble found in \cite{Fujimori:2018nvz} passes through the singular points and the perturbative expansion in small $g=1/k\ll1$ becomes asymptotic, as already found in \cite{Honda:2016vmv}, while for us the perturbative thimble circles around the singularities and the perturbative expansion in $\xi\gg1$, being just a residue calculation, is truncating after finitely many orders\footnote{We thank Masazumi Honda and Tatsuhiro Misumi for useful discussions on these points.}.

We will shortly see that Stokes phenomenon and an asymptotic perturbative expansion are present also in our case although both very different in nature from the analysis of \cite{Fujimori:2018nvz}.
We first focus on the thimble decomposition.

As just discussed, in the present case the following puzzle emerges. It is clear from Figure \ref{spheredecomposition}-(b) that the only non-zero intersection number between the original contour of integration $\Gamma$ (which was running along the real line and circling around the origin in the lower complex $x$ half-plane) and the K thimbles is when we consider the upward manifold associated to the perturbative saddle, i.e. $x_{cr}= -i/(\pi \xi) +O(\xi^{-3})$. 
So in order to compute the path integral from the Picard-Lefschetz decomposition we only need to include the integral over the J cycle that is attached to the perturbative saddle and this contour picks up contributions from all the poles in the upper half plane. Contrary to what usually happens in $2$-d and $4$-d, this includes not only contributions from non-perturbative parts in the same topological sector (vortex-anti-vortex, 2-vortex-2-anti-vortex etc.), but also the contributions from all the other non-perturbative sectors (vortex, anti-vortex, 2-vortex, etc.). For example the second order pole at $x=i$ contains the contributions from the vortex and anti-vortex parts; likewise the pole at $x=2i$ contains the contributions from the 2-vortex, and the 2-antivortex parts, together with the vortex-anti-vortex, and so on.

We would like to have a decomposition that allows us to discern one topological sector from another. 
One might try to move away from the round sphere case, i.e. $b\neq 1$, and indeed if we consider the squashed sphere case we do see the poles splitting. This is easiest seen by looking at the definition of $s_b(x)$ in equation (\ref{eq:sbproduct}). The poles are at $x=imb+in/b$ for $n,m\in\mathbb{Z}$ so for $b\neq 1$ we find first order poles in general, each encoding the contribution coming from a single non-perturbative background. However the Picard-Lefschetz decomposition still has the same problem: we only need to keep the one thimble attached to the saddle corresponding to the perturbative background. Integrating over this J cycle we will pick up all the poles for all the different non-perturbative backgrounds, in every topological sector. 

How do we get a decomposition of the localised path integral in terms of different thimbles, each one associated to a would-be different topological sector hence giving us a manifest resurgence triangle structure?
The solution to this puzzle comes from considering a complexified squashing parameter $b\in\mathbb{C}$ and $|b|=1$, i.e. $b=e^{i\theta}$.  
We will provide a physical interpretation for this complexification in Section \ref{physicalinterpritation} but for the moment let us see what happens to the Picard-Lefschetz decomposition when we consider $b=e^{i\theta}$. 

The first effect is that although the poles are still located at $x=imb+in/b$ they no longer are confined to the positive imaginary axis but form a lattice and the only poles found on the positive imaginary axis are those coming from what will form the trivial topological sector.

To be concrete let us re-examine the case with one chiral and one anti-chiral multiplet.
The partition function (\ref{eq:SquashedChiral}) and effective action, now with general $b$, are given by
\begin{align}
Z^{(1,1)}_{S_b^3}(\xi)&\notag=\int_\Gamma dx e^{2\pi i x \xi}\frac{1}{4\sinh(\pi xb)\sinh(\pi x/b)}\, ,\\
S^{(1,1)}_{eff}(x)&=-2\pi ix\xi +\log\left(2\sinh(\pi xb)\right)+\log\left(2\sinh(\pi x/b )\right)\,.
\end{align}

The singularities are obviously at $mb$ and $n/b$ for $m,n\in\mathbb{Z}$, and while in (\ref{eq:Seff11}) these were second order poles for the partition function we see that now the poles split up and separately carry information about the vortices and the anti-vortices. We notice that in this example there are no contributions from poles with both vortices and anti-vortices, e.g. for example a pole at $b+1/b$. This is very likely because fermion zero modes for these saddles conspire to cancel all their contributions from the path integral. It would be interesting to understand this from the mirror theory.

\begin{figure}[ht]
 
\begin{subfigure}{0.5\textwidth}
\includegraphics[width=0.9\linewidth, height=6cm]{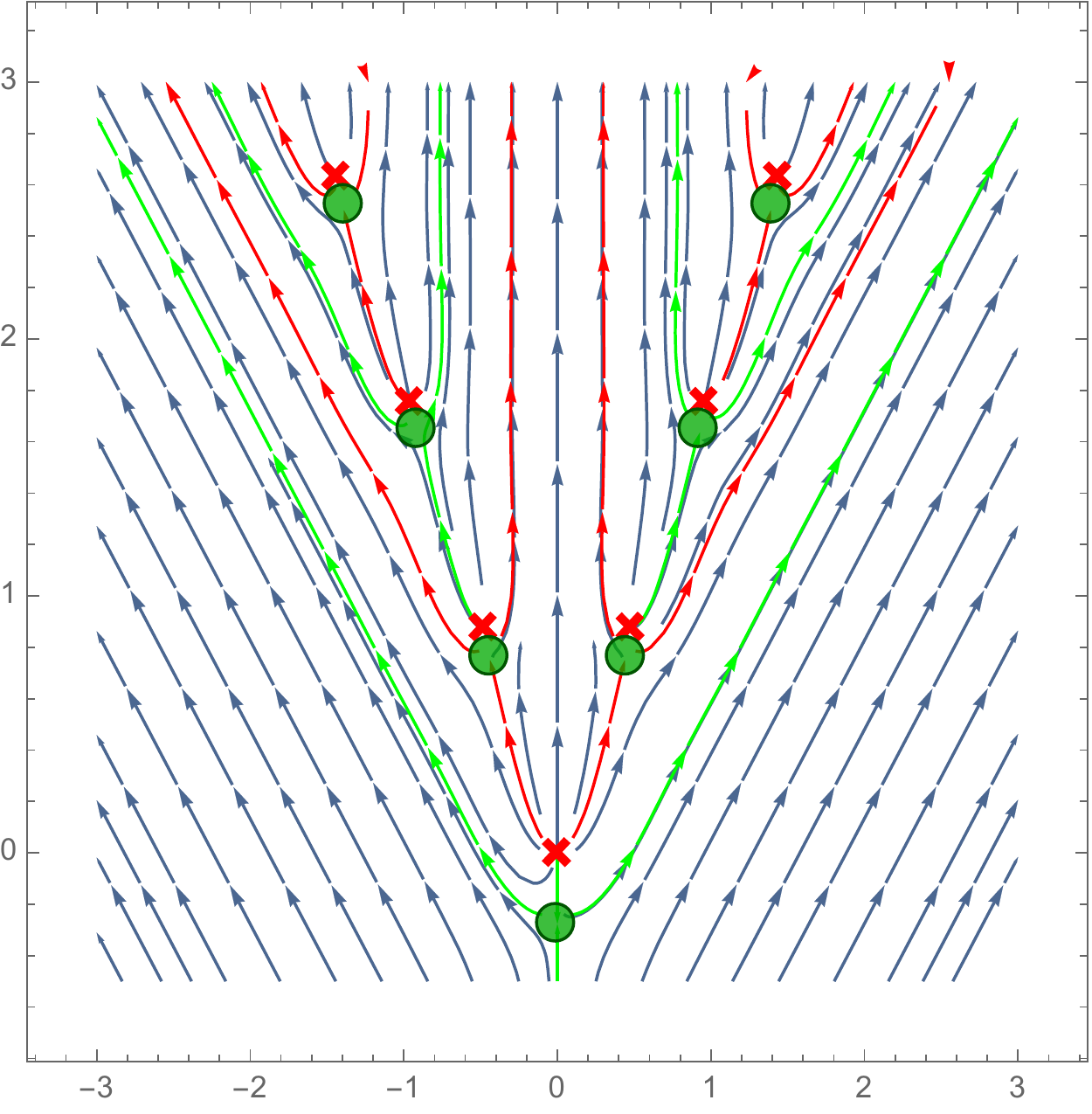} 
\caption{The flow in the upper half plane.}
\label{nonchiralbdecomposition1}
\end{subfigure}
\begin{subfigure}{0.5\textwidth}
\includegraphics[width=0.9\linewidth, height=6cm]{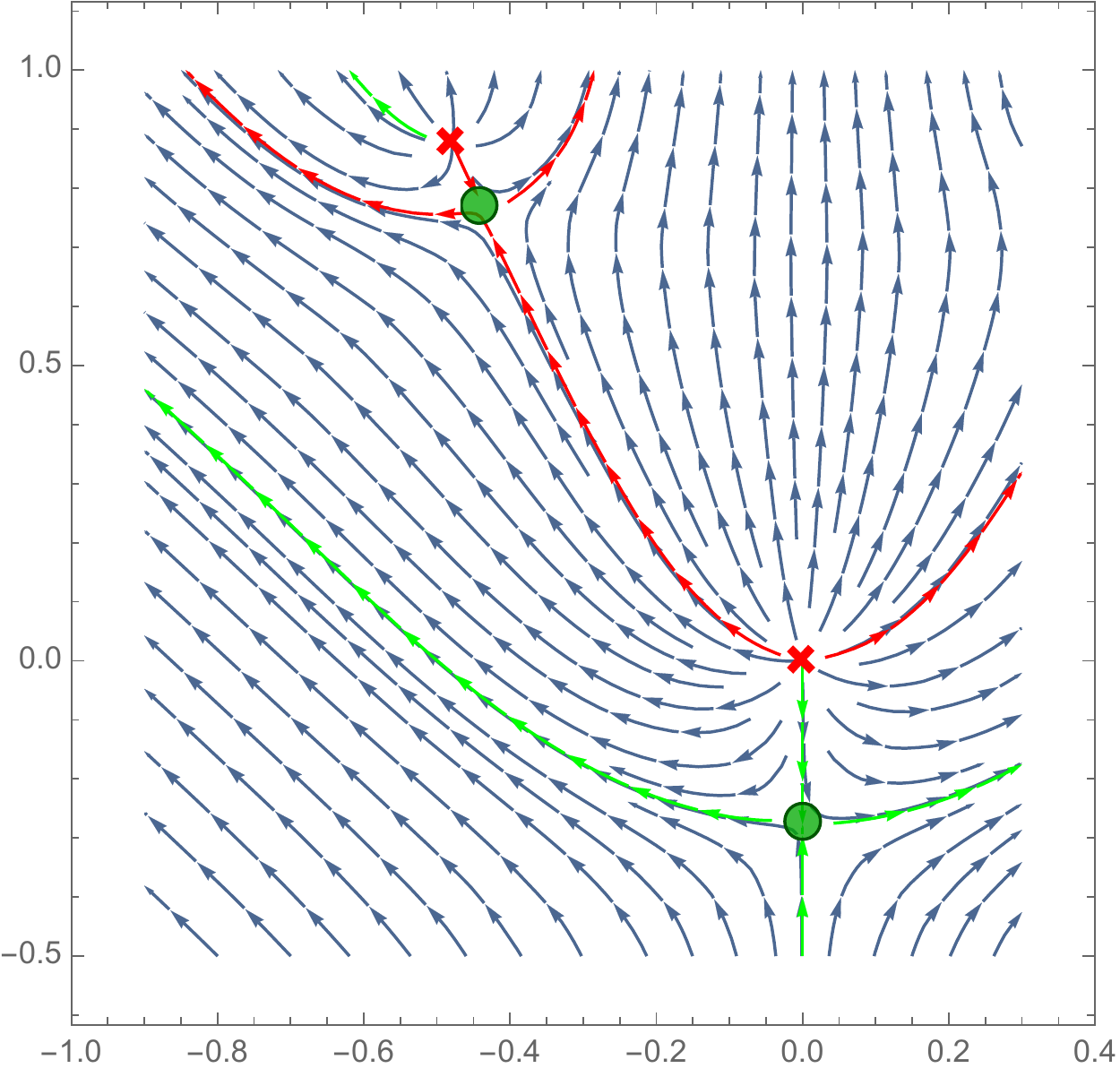}
\caption{Perturbative and 1-vortex saddles.}
\label{nonchiralbdecomposition2}
\end{subfigure}
 
\caption{Morse flow for the theory with one chiral and one anti-chiral, $b=e^{i/2}$, $\xi=1$. The green circles are saddles and the red crosses are poles. From each saddle the downward flows (the J cycles) go off to the sides and eventually off to $\infty$. The upward flows (K cycles) flow up or down to the nearest pole. Only the K cycle from the perturbative saddle hits the real axis.}
\label{nonchiralbdecomposition}
\end{figure}

If we perform a Picard-Lefschetz decomposition as before we obtain Figure \ref{nonchiralbdecomposition}, where we have chosen $b=e^{i/2}$ and $\xi=1$. For this choice of parameters we can easily see the splitting of the poles into contributions from different topological sectors, and as we will shortly discuss in Section \ref{hiddentopologicalanglesection}, complexifying $b$ will effectively introduce a hidden topological angle so we can distinguish between all the sectors with different topological number; for example the vortex sector from the anti-vortex sector. However it is clear from Figure \ref{nonchiralbdecomposition} that with this choice of parameters we still only need to integrate over the $J$ cycle from the perturbative saddle as its K cycle is the only one having non-zero intersection number with the original integration contour $\Gamma$, i.e. we still have not achieved a complete splitting of the path-integral in thimbles for each topological sectors. 

\begin{figure}[ht]
 
\begin{subfigure}{0.5\textwidth}
\includegraphics[width=0.9\linewidth, height=6cm]{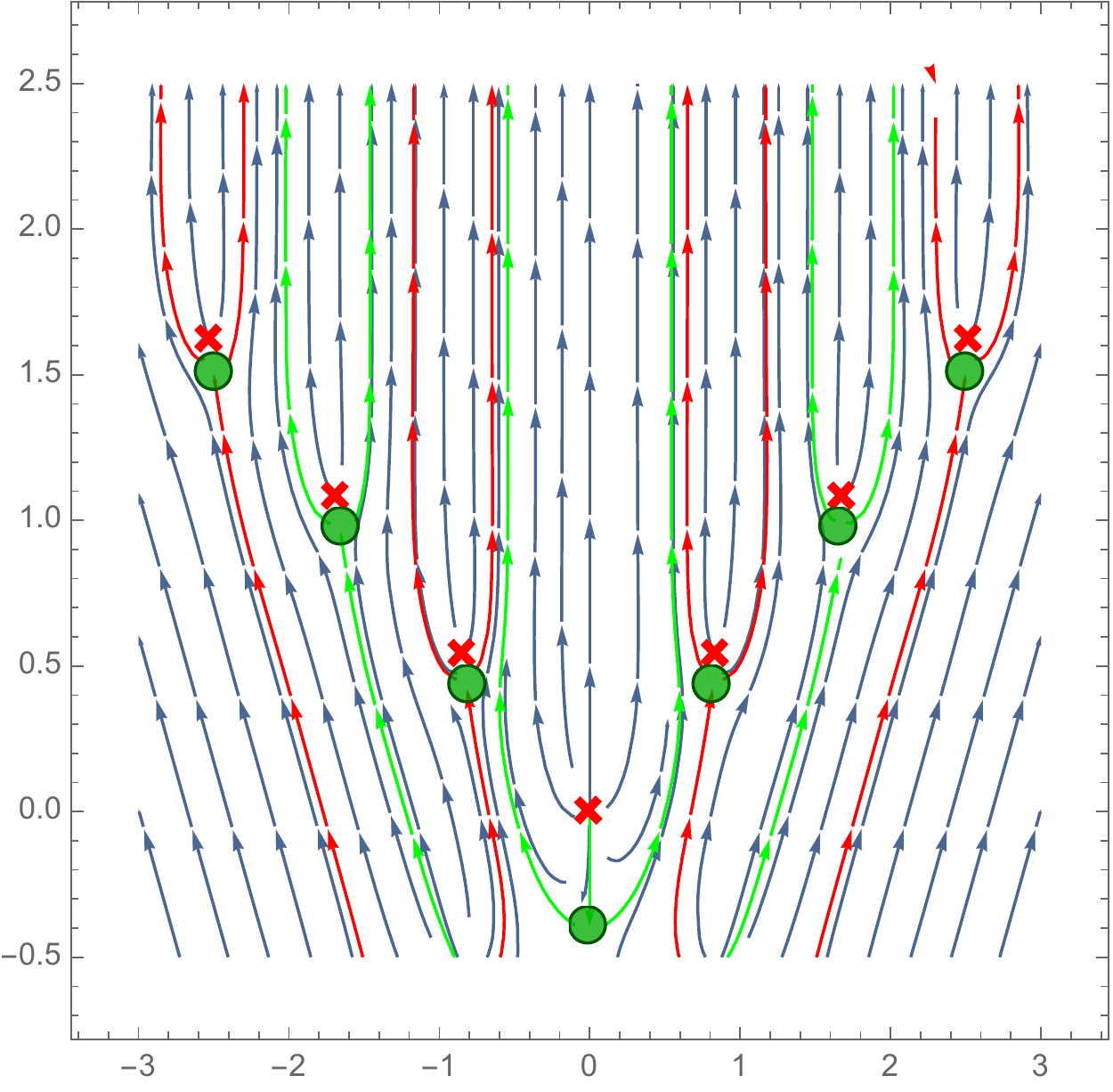} 
\caption{The flow in the upper half plane.}
\label{nonchiralBdecomposition1}
\end{subfigure}
\begin{subfigure}{0.5\textwidth}
\includegraphics[width=0.9\linewidth, height=6cm]{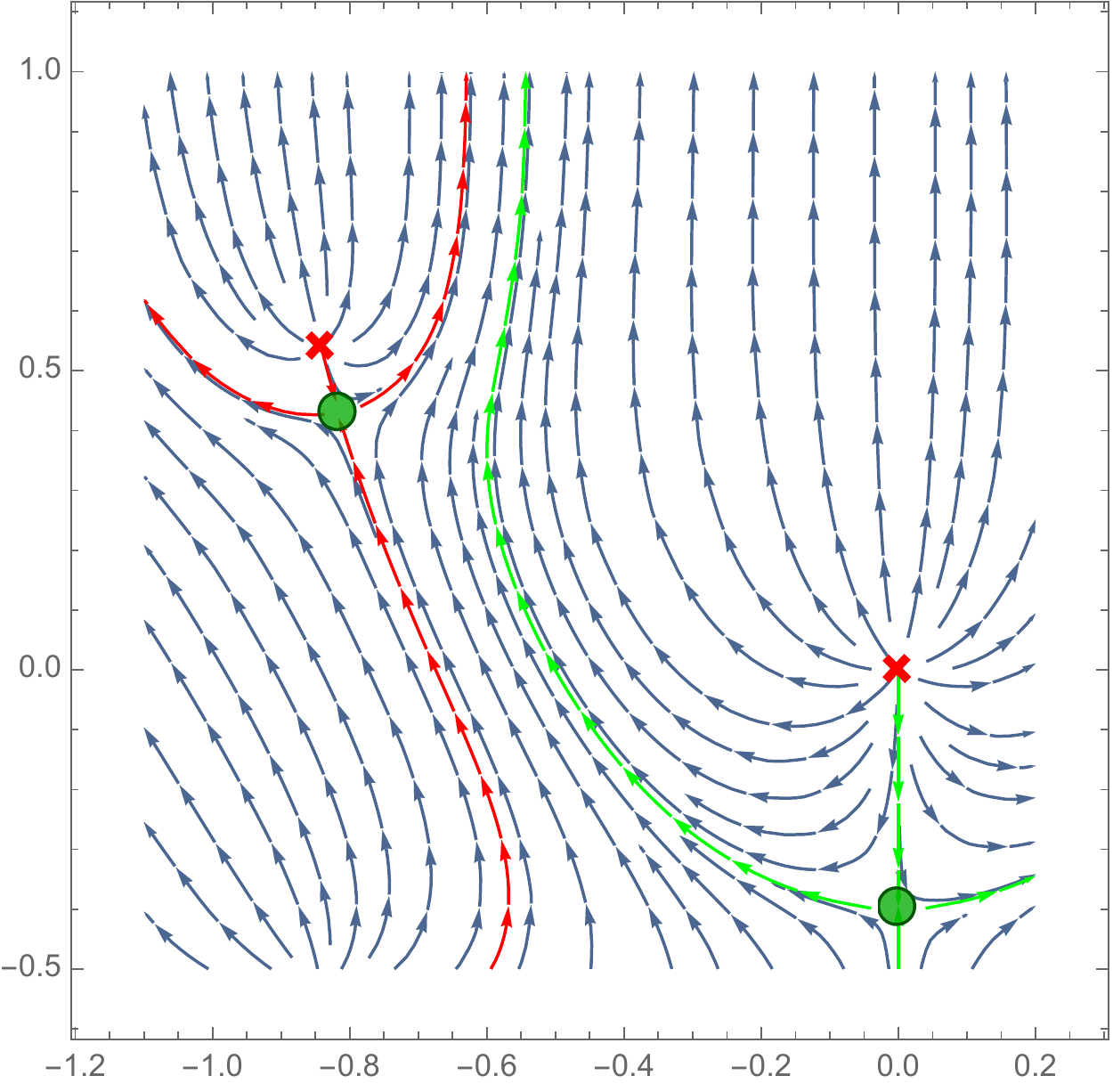}
\caption{Perturbative saddle and 1-vortex saddles.}
\label{nonchiralBdecomposition2}
\end{subfigure}
 
\caption{Morse flow for the theory with one chiral and one anti-chiral, $b=e^{i}$, $\xi=1$. The green circles are saddles and the red crosses are poles. From each saddle the downward flows (the J cycles) go off to the sides and eventually off to $\infty$. The upward flows (K cycles) flow up to the nearest pole and they all intersect the real axis.}
\label{nonchiralBdecomposition}
\end{figure}

The reason for this lies in our choice of parameters $b$ and $\xi$. Let us repeat the Picard-Lefschetz decomposition but this time with $b=e^i$, without changing $\xi=1$, shown in Figure \ref{nonchiralBdecomposition}. All the K cycles intersect the original contour of integration $\Gamma$, hence following our discussion at the beginning of this Section we must include the contributions from the J cycles coming from all the saddles. We moved from the decomposition in Figure \ref{nonchiralbdecomposition} to the one in Figure \ref{nonchiralBdecomposition} by making the argument of $b$ larger. However we could have obtained the same result by cranking up the FI parameter $\xi$. 

As we increase the FI parameter, or alternatively the argument of $b$, more and more K cycles will eventually intersect the original contour $\Gamma$, and hence we have to include in the path-integral more and more J cycles coming from new saddles.
This discontinuous transition is called Stokes phenomenon and its presence is tightly connected with the physical interpretation of the complexification of the squashing parameter. We will expand on this in Section \ref{physicalinterpritation}. Note however that since we are interested in a weak coupling, semi-classical expansion for the path-integral we are actually interested in the limit $\xi\rightarrow\infty$. For this reason, in this limit it is sufficient to include any non-zero complexification of $b= e^{i\theta}$ in order to split the path-integral into the sum of integrals over all of the J cycles in each topological sector as in Figure \ref{nonchiralBdecomposition}.

Let us look at yet another more interesting example given by the theory with two chiral multiplets. The partition function is now
\begin{eqnarray}
Z_{S_b^3}^{(2,0)}(\xi)&=&\int_\Gamma dx \, e^{2\pi i \xi x}\,\left(s_b\left(x+iQ/2\right)\right)^2 \nonumber \\
&=&  \int_\Gamma dx \, e^{2\pi i \xi x-i\pi(x+iQ/2)^2- i \pi \frac{Q^2-2}{12}}\left(\frac{\left(e^{2\pi (bx+ib^2)},e^{2\pi ib^2}\right)_\infty}{\left(e^{2\pi x/b},e^{-2\pi i/b^2}\right)_\infty}\right)^2\;,
\end{eqnarray}
giving us the effective action
\begin{eqnarray}
S_{eff}^{(2,0)}(x)&=&-2\pi ix\xi+i\pi\left(x+\frac{iQ}{2}\right)^2 + i \pi\frac{Q^2-2}{12}-2\log\left[ \left(e^{2\pi (bx+ib^2)},e^{2\pi ib^2}\right)_\infty\right]\nonumber \\
&&+2\log\left[ \left(e^{2\pi x/b},e^{-2\pi i/b^2}\right)_\infty\right] \;\;\; .
\end{eqnarray}
Note that the q-Pochhammer $(a;q)_\infty$ is only defined when the modulus of the second argument is less than one. Thus for the above expression for $S_{eff}$ to make sense we must have $e^{2\pi ib^2}$ and $e^{-2\pi i/b^2}$ both with modulus less than one. Thus we will only consider the case where $b=e^{i\theta}$ for $0<\theta<\pi/2$ (or alternatively $-\pi <\theta<-\pi/2$). We will be easily able to relate this to the case $-\pi/2<\theta<0$ (respectively $\pi/2 <\theta<\pi$) by the vortex $\leftrightarrow$ anti-vortex symmetry, i.e. $b\to b^{-1}$.

\begin{figure}[ht]
 
\begin{subfigure}{0.5\textwidth}
\includegraphics[width=0.9\linewidth, height=6cm]{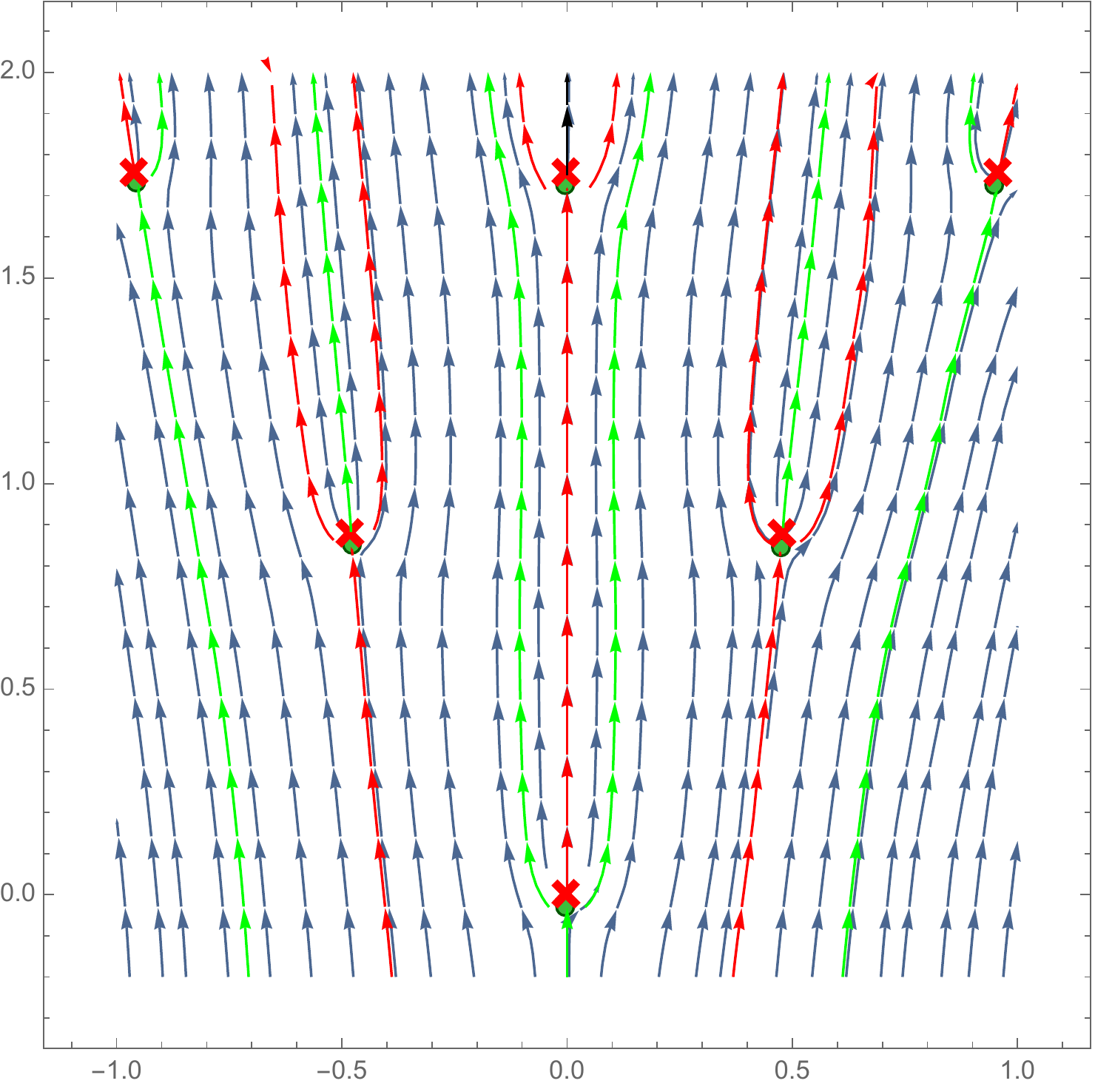} 
\caption{The flow in the upper half plane.}
\label{chiralBdecomposition1}
\end{subfigure}
\begin{subfigure}{0.5\textwidth}
\includegraphics[width=0.9\linewidth, height=6cm]{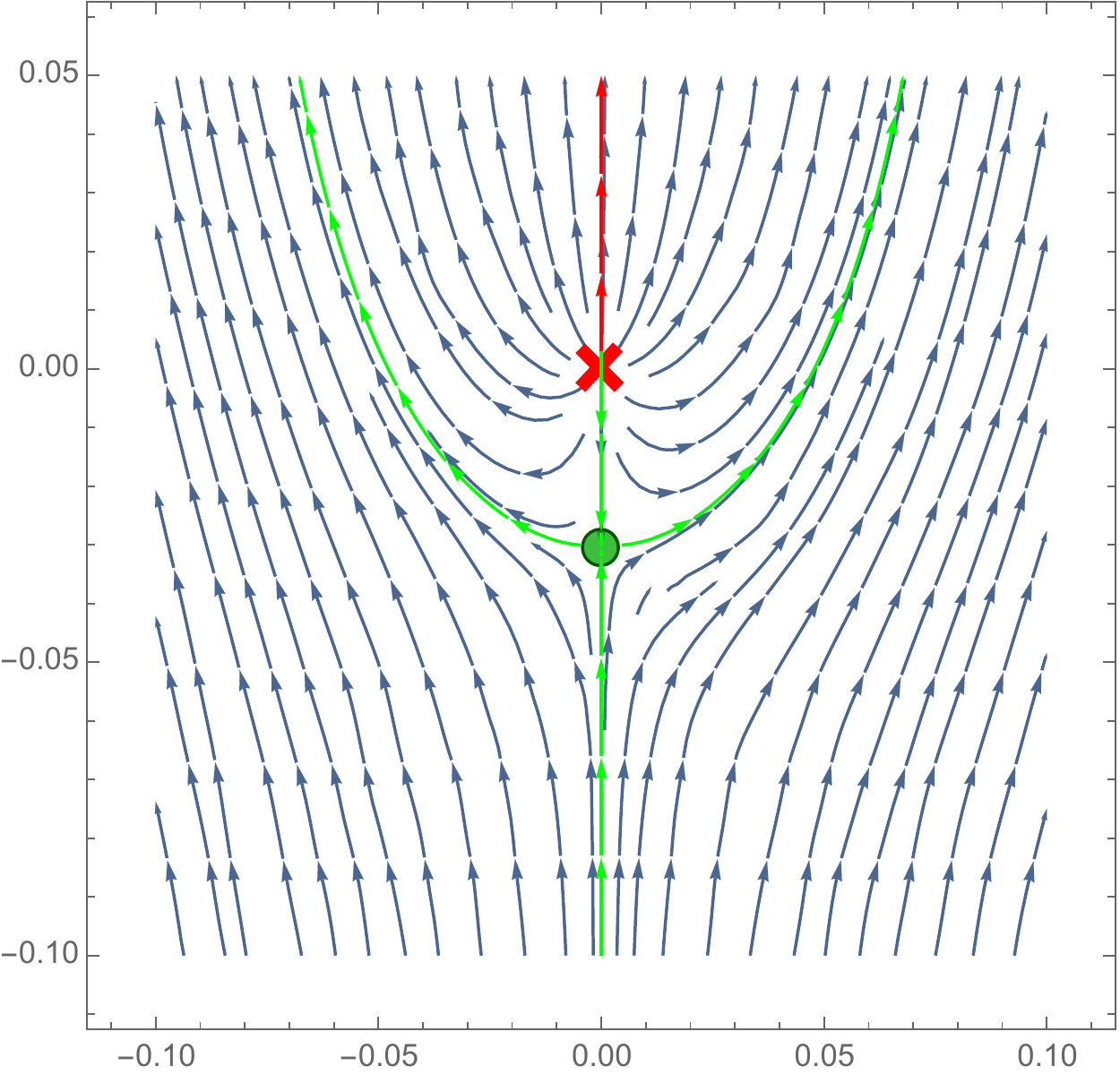}
\caption{Perturbative saddle and its nearest pole.}
\label{chiralBdecomposition2}
\end{subfigure}
 
\caption{Morse flow for the theory with two chirals, $b=e^{i/2}$, $\xi=12$. The green circles are saddles and the red crosses are poles. From each saddle the downward flows (the J cycles) go off to the sides and eventually off to $\infty$. The upward flows (K cycles) flow up to the nearest pole, and down to either the nearest pole, or past the real axis if they are the lowest saddle in their given topological sector.}
\label{chiralBdecomposition}
\end{figure}

We now perform the Picard-Lefschetz decomposition as above, which is shown in Figure \ref{chiralBdecomposition}. As it is manifest from Figure \ref{chiralBdecomposition} when we decompose the path-integral we need to include all of the J cycles coming from the lowest saddle point in each topological sector. The main novelty in this example is that now we do have contributions coming for all the non-perturbative solutions, i.e. we get contributions from $m$-vortex-$n$-anti-vortex saddles for $m,n\in\mathbb{N}$. We do not need to include all of their J cycles, we just need the J thimble coming from the lowest (real part of the) action solution in each topological sector. For example integrating over the J cycle from the perturbative saddle will pick up the contributions from all the saddles in the trivial topological sector, i.e. all the k-vortex-k-anti-vortex saddles. We have just recovered the full resurgence triangle structure.

\subsection{Recovering the resurgence triangle}\label{hiddentopologicalanglesection}
In this Section we kept on referring to the critical points of the effective action as different topological sectors despite our $3$-dimensional theory not having a topological theta angle characterising the usual $4$-d decomposition of the path-integral into different instantonic sectors.
The reason for our ``abuse'' of terminology lies in the complexification of the squashing parameter and the subsequent appearance of what seems to be very similar to a topological angle.

Let us go back to the general partition function (\ref{localisedpartition}) and look more closely at the poles of the one-loop determinant. Here the poles lie at $x=imb +in/b$ for $m,n\in\mathbb{N}$ and the classical action term in the integrand, $e^{2\pi i\xi x}=e^{-S_c}$, evaluated at these locations is $e^{2\pi i\xi (imb+in/b)}$.
When $b = e^{i\theta}$ we can define
\begin{equation}
Q= b+\frac{1}{b}=2\cos\theta\,,\qquad\qquad \Theta = -i \left(b-\frac{1}{b}\right) = 2\sin\theta\,.\label{eq:Theta}
\end{equation}
Now we see that the classical action evaluated at each of the poles can suggestively be rewritten as
\begin{align}
S_c(m,n)&\notag=2\pi i\xi (imb+in/b) = \pi\xi[ (m+n) Q+i(m-n)\Theta]\\
& = \pi \xi \left( \vert N \vert Q + i \Theta N\right)+ 2\pi \xi Q k\,,\label{eq:action}
\end{align}
where $N=m-n$ and $k = \mbox{min}(m,n)$.
In terms of these new variables $(N,k)$ it is now clear that $S_c(m,n)$ corresponds to the $k$ vortex-anti-vortex solution on top of the $N$-vortex topological sector (anti-vortex sector if $N<0$). The case $N=0$, i.e. $m=n$, is then related to the topologically trivial sector, directly connected to the usual perturbative vacuum.

Importantly we notice that the classical actions of these solutions are now complex: the imaginary part of the action is related to a hidden topological angle (HTA) $\Theta$. When $b$ is real $\Theta$ vanishes and we cannot decompose the path-integral into different topological sectors but as soon as we complexify $b$, $\Theta$ becomes non-zero and the HTA allows us to identify a column of non-perturbative contributions topological sector by topological sector. This is reminiscent of the theories studied in \cite{Behtash:2015kna,Behtash:2015kva,Behtash:2015zha,Behtash:2015loa,Dunne:2016jsr}.

Note however a key difference: for theories with a genuine topological angle the action of non-perturbative objects, being that for example instantons in $4$-d or vortices in $2$-d, takes the schematic form $S =\vert N \vert/g + i \Theta N$ for some coupling constant $g$ and topological number $N$. In particular the real and imaginary part of the on-shell actions are not correlated, i.e. the $\theta$ angle has nothing to do with the coupling constant. In the present case however we have that both the real and imaginary part of the saddles action (\ref{eq:action}) depend from the coupling $\xi$, this will have important repercussions on the resurgent structure of the theory.

Forgetting this issue for the moment we can thus split the partition function into a sum over topological sectors in a transseries:
\begin{eqnarray}\label{toposplit}
Z_{S_b^3}^{(N_c,N_a)}(\xi)=\sum\limits_{N=-\infty}^\infty e^{-\pi\xi Q \vert N \vert + i \pi \xi \Theta N}\zeta_N(\xi)\;,
\end{eqnarray}
where $\zeta_N(\xi)$ contains the contributions from all the $k$ vortex-anti-vortex saddles in the $N^{th}$ topological sector.

The function $\zeta_N(\xi)$ precisely corresponds to the $N^{th}$ column of the resurgence triangle presented in Figure \ref{Resurgencetriangle}
\begin{equation}\label{eq:generalTS}
\zeta_N(\xi) = \sum_{k=0}^\infty e^{-2\pi \xi Q k} \Phi_N^{(k)}(\xi)\,,
\end{equation}
a sum of perturbative expansions, $\Phi_N^{(k)}(\xi)$, on top of a $k$ vortex-anti-vortex background in the $N^{th}$ topological sector.
In Section \ref{resurgenceanalysis} we will show how one can use resurgent theory to extract from just one of the $\Phi_N^{(k)}(\xi)$ all the other  $\Phi_N^{(k')}(\xi)$ belonging to the same topological sector.

\begin{figure}[ht]
\begin{center}
\begin{subfigure}{0.35\textwidth}
\includegraphics[width=0.9\linewidth, height=4cm]{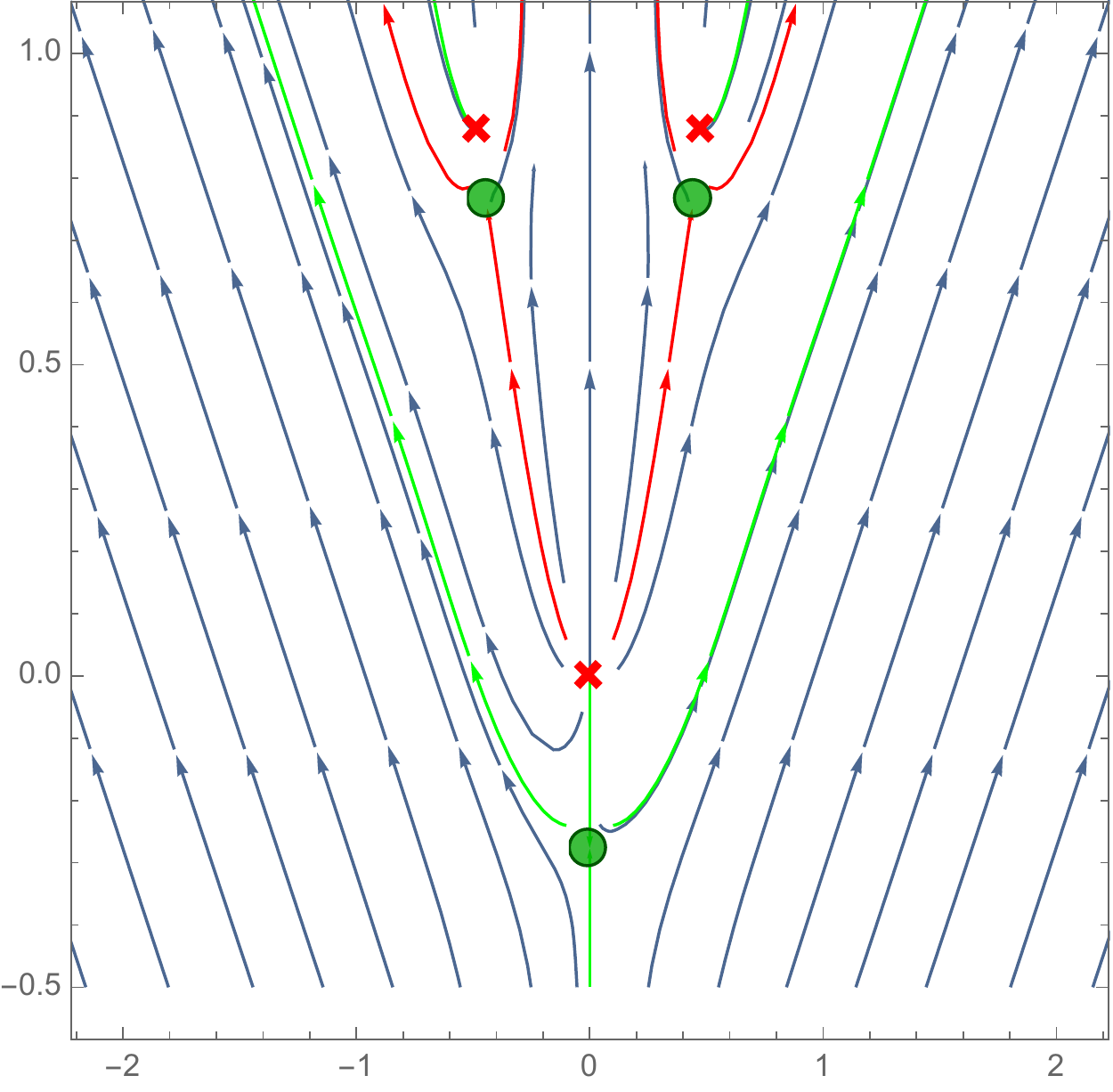} 
\caption{$b=e^{i/2}$}
\label{Stokes1}
\end{subfigure}
\begin{subfigure}{0.30\textwidth}
\includegraphics[width=0.9\linewidth, height=4cm]{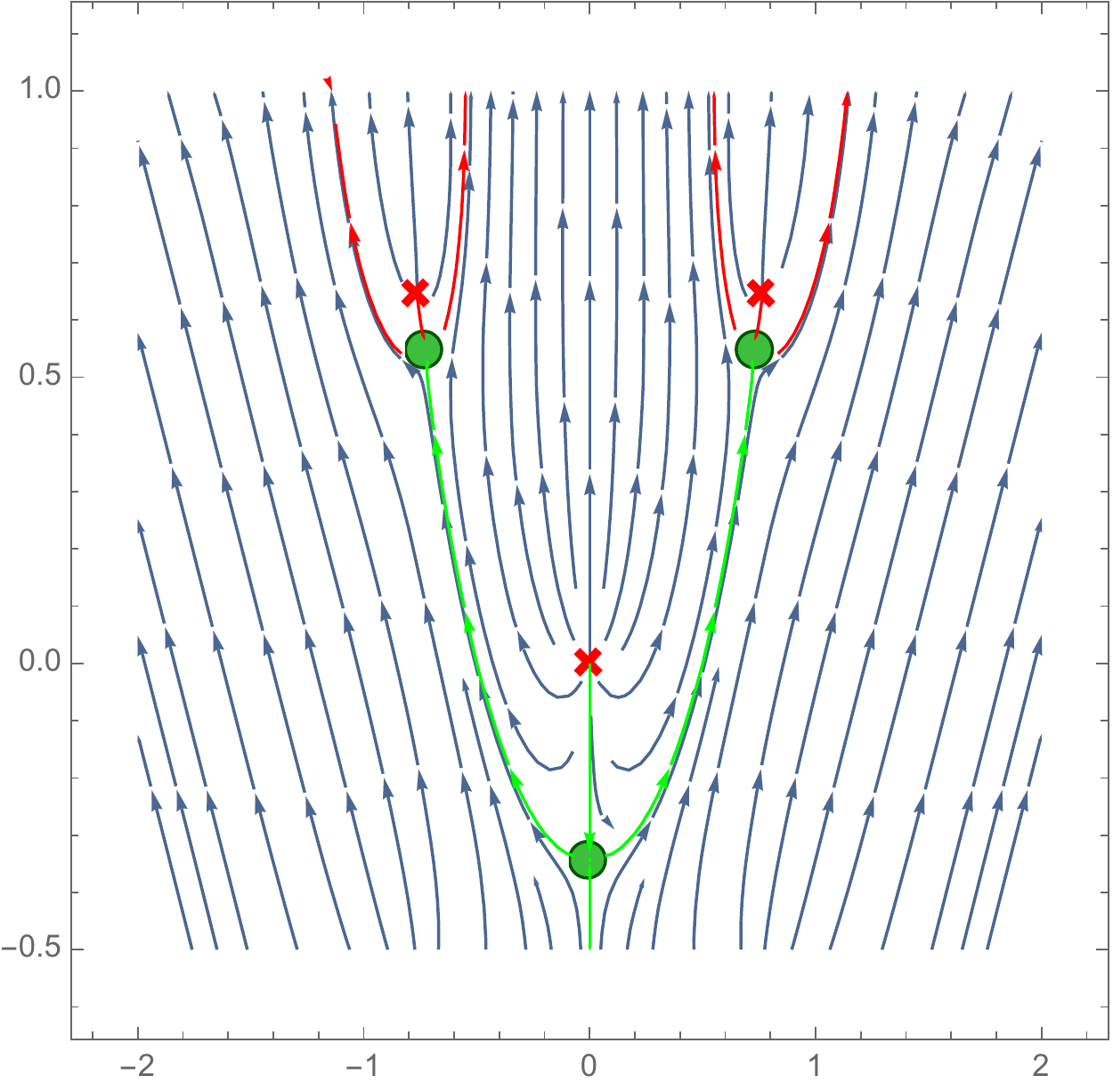}
\caption{$b= e^{0.87i}$, at stoke line.}
\label{Stokes2}
\end{subfigure}
\begin{subfigure}{0.30\textwidth}
\includegraphics[width=0.9\linewidth, height=4cm]{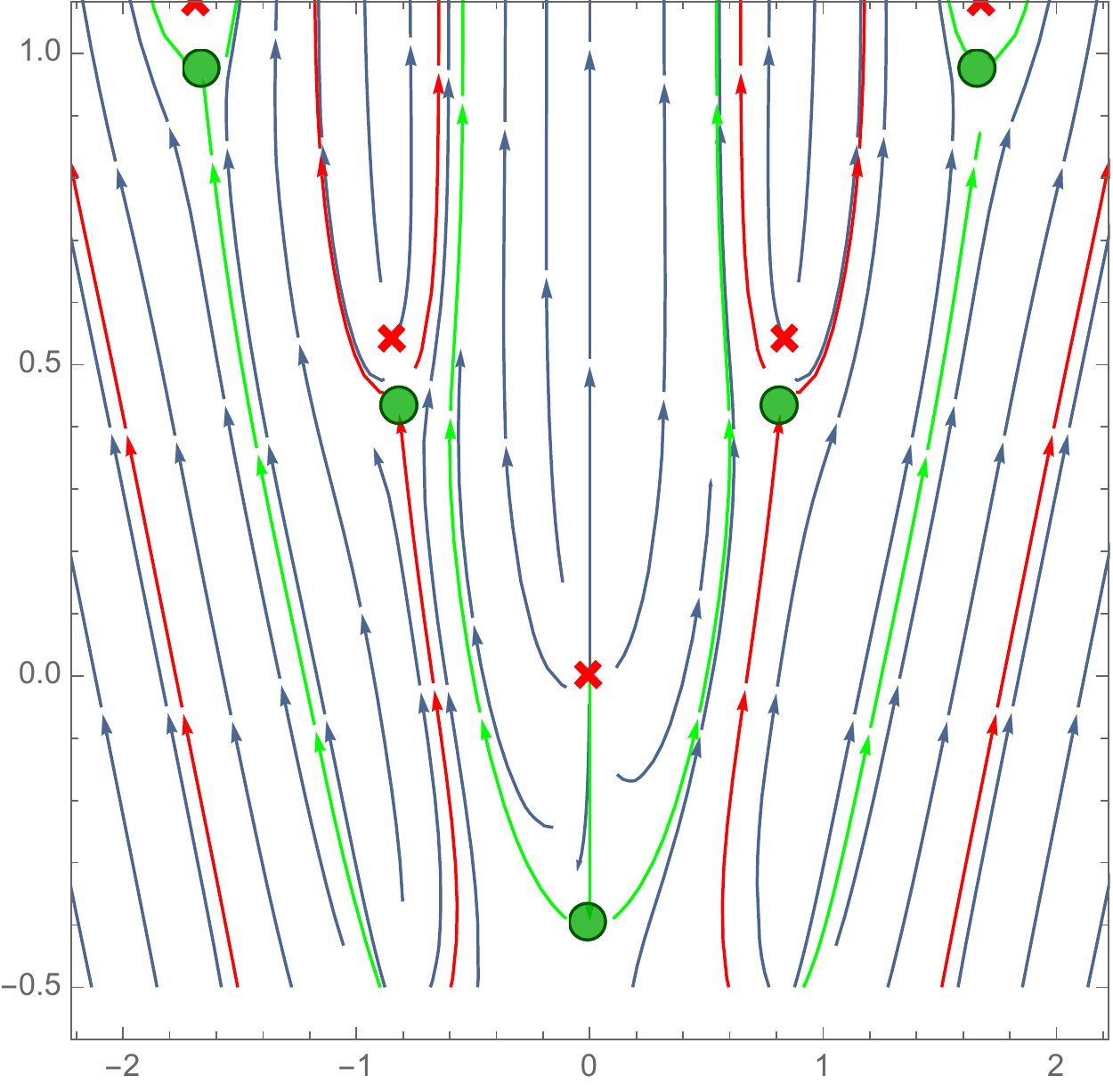}
\caption{$b=e^{i}$}
\label{Stokes3}
\end{subfigure}
 
\caption{Morse flow for the theory with one chiral and one anti-chiral for $b=e^{i/2}$, $b= e^{0.87i}$ and $b=e^i$ all with $\xi=1$. The green circles are saddles and the red crosses are poles. As $\theta$ increases the position of the saddles and the J and K cycles change. Figure (b) shows when the Stokes crossing happens. At this point the J cycle from the perturbative saddle connects with the K cycle from the vortex and anti-vortex. As $\theta$ increases beyond this value, Figure (c), the K cycles from vortex/ anti-vortex no longer flow to the perturbative pole, but crosses through the real axis.}
\label{stokesdecomposition}
\end{center}
\end{figure}

\subsection{Complexified squashing and Stokes phenomenon}\label{physicalinterpritation}

We would like to understand now the physical interpretation of this complexified squashing parameter.
In \cite{Closset:2012ru} (see also \cite{Alday:2013lba}) the authors studied the rigid limit of $3$-d new minimal supergravity to find all possible backgrounds (metric and auxiliary fields) admitting rigid supercharges.
In particular for theories with four supercharges and for which the three dimensional manifold is an $S^1$ fibration over $S^2$ the metric takes the form 
\begin{equation}\label{eq:metric}
ds^2 =h^2 (d\psi +2 \sin^2 \frac{\theta}{2} d\phi)^2 +(d\theta^2 +\sin^2\theta d\phi^2)\,,
\end{equation}
where $(\theta,\phi)$ are the usual coordinates on $S^2$, $\psi$ is the angular coordinate parametrising the $S^1$ Hopf fibre over $S^2$ and $h\in\mathbb{R}\setminus\{0\}$ that we can parametrise as $ h = (b+1/b)/2$.
However to have rigid supersymmetry we must also turn on some background fields, in particular a vector field $V^\mu$ must be present and it takes the form
\begin{equation}
V^\mu \partial_\mu = \frac{b-b^{-1}}{b+b^{-1}} \,\partial_\psi\,,
\end{equation}
hence a never vanishing Killing vector associated to the $U(1)$ isometry of the $S^1$ fibre.
Note that this supersymmetric background has actually two branches.
When $b>0$ both $h$ and $V_\mu$ are real and this corresponds precisely to the squashed $S_b^3$ case discussed so far. However we can also pick $b=e^{i\theta}$ for some $\theta \in [0,2\pi]$ (not to be confused with one of the coordinates of the $S^2$). The metric is still completely real however the Killing vector has now become purely imaginary.

We can understand this complex squashing as turning on a chemical potential for the $U(1)$ rotation or equivalently, thanks to the non-trivial fibering (\ref{eq:metric}), for the $J_z$ rotation of $S^2$. The branch $b>0$ corresponding to real squashing is continuously connected to the branch $b=e^{i\theta}$ corresponding to the introduction of an omega-deformation, effectively rotating the $S^2$ along its axis. 
When this chemical potential is turned on we have that vortices will become weighted by $(b-1/b)/(b+1/b)= i\Theta/Q$ while anti-vortices will be weighted by $-(b-1/b)/(b+1/b)= -i\Theta/Q$ exactly has shown in equation (\ref{eq:action}).
For real $b$ we cannot distinguish between different topological sectors via Picard-Lefschetz decomposition, but the moment we include a phase in $b$ the topological sectors split and we can distinguish between them in our Picard-Lefschetz decomposition.

Furthermore we can also understand the reason for the appearance of Stokes phenomenon as we vary the argument of $b$ at fixed FI $\xi$, or similarly modifying the value of the FI parameter for fixed, non-zero argument of $b$.
The reason is that the FI parameter is regulating the size of the vortices; for infinite FI parameter the vortices are point-like objects perfectly localised at the north and south poles of the $S^2$. 
On the other hand for finite FI parameter vortices have a size and they are not perfectly localised at the poles, and have some overlap at the equator.

There is now some play off between the FI parameter and the phase of $b$. If the value of $\xi$ is not large enough we cannot immediately distinguish between topological sectors via Picard-Lefschetz decomposition the moment we switch on a phase for $b$. For a given phase of $b$, the FI parameter needs to be sufficiently large, i.e. the vortices need to be sufficiently localised at the north and south poles, before we can distinguish between sectors. 
For small argument of $b$ the imaginary part of the action at the perturbative saddle of the effective action (the saddle just below the pole at the origin) is small but non-zero, and it will generically be different from the imaginary part of the action at the non-perturbative saddles of the effective action. As the argument of $b$ (or the FI) increases, so does the imaginary part of the action of the saddles in the non-perturbative sectors. At some point the imaginary part of the classical action of the perturbative saddle will equal the vortex and the anti-vortex one and it will be possible to construct a thimble joining these different saddles, i.e. we will be at a Stokes line, see Figure \ref{stokesdecomposition} - (b).  At this point the J cycle from the perturbative saddle hits the vortex and anti-vortex saddles. Increasing $b$ even more and we will cross this Stokes line, the J cycle jumps over the saddle from the vortex saddle, and in our decomposition we now have to include the J cycles from the vortex and the anti-vortex as well, \ref{stokesdecomposition} - (c).

It should be in principle possible to derive our analysis as the limit of vanishing Chern-Simons level, i.e. strong coupling $g=1/k\to\infty$, and vanishing real masses of the thimble decomposition carried out in \cite{Fujimori:2018nvz}.  However it is very likely that this is a singular limit since the tails of the thimbles, i.e. the relative homology of good regions ($\mbox{Re} \,S_{eff}(x)>0$) in the complex $x$-plane, change discontinuously for $g>0$ and $g=0$.
It would also be interesting to analyse more in details the monodromy structure of these thimbles for intermediate values of $\xi$ and understand the connection between these Stokes jumps and the analysis carried out in \cite{Beem:2012mb}. 

Furthermore it was observed in \cite{Hosomichi:2010vh} that the building blocks (\ref{eq:sbproduct}) to compute the $3$-d $\mathcal{N}=2$ partition functions on a round sphere, i.e. $b=1$, are directly related to the structure constants in $2$-d Liouville with central charge $c=25$. Roughly speaking our $3$-d theory is realised on the domain wall of two $S$-dual $\mathcal{N}=4,\,4$-d gauge theories which are in turn related to $2$-d Liouville via AGT correspondence. 
Subsequently in \cite{Hama:2011ea} this correspondence was generalised to the $3$-d squashed sphere case, i.e. $b>1$, and the structure constants of Liouville with central charge $c=1+6Q^2=1+6(b+b^{-1})^2$.

Our complexification of the squashing parameter would now allow us to interpolate continuously between ``standard'' space-like Liouville, for which $b\in\mathbb{R}$ and $c= 1+6(b+b^{-1})^2\geq 25$, and time-like Liouville, for which $b=i\hat{b}$ with $\hat{b}\in\mathbb{R}$ $c=1-6(\hat{b}-\hat{b}^{-1})^2\leq 1$. We just need to use $b=e^{i\theta}$ with $\theta\in[0,\pi/2]$ to connect $b=1$ to $\hat{b}=1$. 
It would be extremely interesting to follow the analytic continuation of the integration contours of the path-integral for Liouville along this path in the complex $b$ plane following the works \cite{Harlow:2011ny,Bautista:2019jau}.

\section{Resurgence analysis}\label{resurgenceanalysis}

Now that we have understood how to decompose the localised path-integral as a sum over thimbles each one of them associated to a different topological sector we want to analyse whether or not in each topological sector one can retrieve higher non-perturbative corrections from the purely perturbative data by means of resurgent analysis. We will shortly see that it will be necessary to introduce a Cheshire Cat deformation to make this resurgent structure manifest. However we will first start our discussion with the undeformed theory to clarify the necessity of this deformation.

\subsection{Undeformed theory}

\begin{figure}[ht]
\centering
\includegraphics[width=0.5\textwidth]{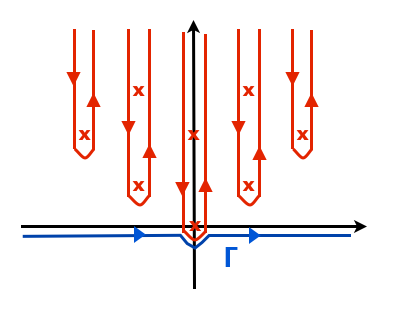}
\caption{Picard-Lefschetz decomposition of the original contour of integration $\Gamma$. Each thimble identifies a different topological sector.}
\label{picardLefschetz}
\end{figure}

Let us analyse more in detail, and thimble by thimble, the analytic structure of the localised path-integral and for concreteness we will focus to the case of two chirals although it is easy to repeat the analysis in theories with any other matter content. As argued in the previous Section we can decompose the path integral into contours as shown in Figure \ref{picardLefschetz}.

Thus we can write the path integral as
\begin{eqnarray}
Z_{S_b^3}^{(2,0)}(\xi)&=&\int_\Gamma dx\; e^{2\pi i\xi x} \left(s_b(x+iQ/2)\right)^2 \nonumber \\
&=&\sum_{n\in\mathbb{Z}}\, \int_{\Gamma_n} dx\; e^{2\pi i\xi x} \left(s_b(x+iQ/2)\right)^2\; ,
\end{eqnarray}
where $\Gamma_N$ is the contour associated to the $N^{th}$ topological sector, $N\leq0$ being the $-N$ vortex sector, while $N>0$ being the $N$ anti-vortex sector, as schematically depicted in Figure \ref{picardLefschetz}.

The contour $\Gamma_{-N}$, for $N\geq 0$, runs vertically starting from $+i\infty +\mbox{Re} ( i N b) -\epsilon$ circles around the point $i N b $ and goes back to $+ i\infty +\mbox{Re} (i N b) +\epsilon$. Similarly the contour $\Gamma_{N}$, with $N>0$,  runs vertically starting from $+i\infty +\mbox{Re} (i N/b) -\epsilon$ circles around the point $i N/ b $ and goes back to $+ i\infty +\mbox{Re} (i N/b) +\epsilon$.
The first pole in each topological sector is to be found at $iNb$ or $iN/b$ for the contour $\Gamma_{-N}$ or $\Gamma_N$ respectively. For each one of these integrals we can shift the integration variable to move the first pole in its topological sector to the origin, namely we rewrite the contours as $\Gamma_{-N} = i N b +\Gamma_0$  or $\Gamma_N = i N/b +\Gamma_0$, for $N\geq0$. This shift in integration variable will bring out an explicit exponential of the classical action factor as the weight of each topological sector. 
The partition function is then
\begin{eqnarray}
Z_{S_b^3}^{(2,0)}(\xi)&=&\int_{\Gamma_0} dx \;e^{2\pi i\xi x} \left(s_b(x+iQ/2)\right)^2 \nonumber \\
&&+ \sum\limits_{N>0}e^{-2\pi\xi Nb}\int_{\Gamma_0} dx \;e^{2\pi i\xi x} \left(s_b(x+iNb+iQ/2)\right)^2\nonumber \\
&&+ \sum\limits_{N>0}e^{-2\pi\xi N/b}\int_{\Gamma_0} dx \;e^{2\pi i\xi x} \left(s_b(x+iN/b+iQ/2)\right)^2 \nonumber \\
&=&\nonumber\zeta_0(\xi,b)+\sum\limits_{N>0}e^{-2\pi\xi Nb}\zeta_N(\xi,b) +\sum\limits_{N<0}e^{2\pi\xi N/b}\zeta_N(\xi,b) \;\;\;\\
&=&\label{eq:TopoTS}\sum\limits_{N=-\infty}^\infty e^{-\pi\xi Q \vert N \vert + i \pi \xi \Theta N}\zeta_N(\xi,b)\,,
\end{eqnarray}
where again $Q= b+1/b$ while $\Theta = -i (b-1/b)$.
Upon complexification of the squashing parameter we have that the Picard-Lesfchetz decomposition of the path-integral can be directly seen as a manifestation of the resurgence triangle precisely as presented in equation (\ref{toposplit}) and the related discussion. 

Now let us zoom in on the topologically trivial sector, i.e. $N=0$, and analyse its corresponding contour integral. We have
\begin{eqnarray}
\zeta_0(\xi,b)&=&\int_{\Gamma_0} dx\; e^{2\pi i x \xi}s_b\left(x+iQ/2\right)^2 \nonumber \\
&=&\int_{\Gamma_0} dx \;e^{2\pi i x \xi} \prod\limits_{n,m\geq 0}^{\infty} \frac{((m+1)b+(n+1)/b-ix)^2}{(mb+n/b+ix)^2} \nonumber \\
&=&\label{eq:zerosec}\int_{\Gamma_0} dx\; e^{2\pi i x \xi}\prod\limits_{m=0}^{\infty}\frac{((m+1)(b+1/b)-ix)^2}{(m(b+1/b)+ix)^2}H_0(x)\; ,
\end{eqnarray}
where we define $H_0(x)$ by
\begin{eqnarray}\label{eq:H0}
H_0(x)=\prod_{m\neq n}\frac{((m+1)b+(n+1)/b-ix)^2}{(mb+n/b+ix)^2}\;\;\; .
\end{eqnarray}

$H_0(x)$ can be regularised using q-Pochhammers, but the important property is that it is entire along the contour $\Gamma_0$ as well as in the region define by its interior. 
On the other hand, the remaining infinite product in the integral can be regularised (see for example the Appendix of \cite{Dorigoni:2017smz}) to give
\begin{eqnarray}
\zeta_0(\xi,b)&=&\int_{\Gamma_0} dx\; e^{2\pi i x \xi}\frac{\Gamma\left(\frac{ix}{Q}\right)^2}{\Gamma\left(1-\frac{ix}{Q}\right)^2}H_0(x)\;\;\; .
\end{eqnarray}
Note that if we were to replace $H_0(x)\to 1$ we would obtain precisely the contribution from the topologically trivial sector to the partition function of the $\mathcal{N}=(2,2)$ $\mathbb{CP}^{1}$ model on $S^2$ discussed in \cite{Dorigoni:2017smz} where the chiral fields have effective charge $q=1/Q$. Perhaps not surprisingly this function $H_0(x)$ is storing all the information regarding the additional $S^1$ and all the different topological sectors. 

This integral can be performed by summing over the residues of the poles on the positive imaginary axis and the answer we get is of the form
\begin{eqnarray}\label{eq:TSundef}
\zeta_0(\xi,b)=\sum\limits_{k=0}^\infty e^{-2\pi \xi k Q}\zeta_{0,k}(\xi,b) \;\;\; .
\end{eqnarray}
We have denoted by $\zeta_{N,k}(\xi,b)$ the contribution from the $k$ vortex-anti-vortex saddle on top of the $N^{th}$ topological sector. For example we have
\begin{eqnarray}\label{undeformedcoefficients}
\zeta_{0,0}(\xi,b)&=&(2 \pi Q)^2 \xi H_0(0) \left(1+\frac{i H_0'(0)}{H_0(0)}(2\pi\xi)^{-1}-\frac{4\gamma}{ Q } (2\pi\xi)^{-1}\right)\,, \nonumber \\
\zeta_{0,1}(\xi,b)&=&(2 \pi  Q)^2 \xi H_0(i Q) \left( 1 -\frac{i  H_0'(i Q)}{H_0(iQ)}(2\pi \xi)^{-1} + \frac{4(1- \gamma)}{Q}(2\pi\xi)^{-1}\right)\;\;\; .
\end{eqnarray} 
The values of the function $H_0$ and its derivatives at these special points can be computed making use of the functional relations (\ref{eq:functional2}) and the known residues for $s_b$, for example from (\ref{eq:Residue}) one can easily see that $H_0(0) = 1/(2\pi Q)^2$. The actual values will not play any role in what follows so we will keep them in this implicit form.

Note importantly that these are the perturbative expansions around each of the classical non-perturbative backgrounds, and they are not asymptotic series in $\xi$, but in fact they truncate after finitely many orders. Thus at first sight it looks like we cannot apply resurgence analysis to this theory.

We can of course repeat this analysis in all of the topological sectors. For the $N^{th}$ topological sector we find
\begin{eqnarray}
\zeta_N(\xi,b)&=&\label{eq:TopSecIntegral}\int\limits_{\Gamma_0} dx\, e^{2\pi i x \xi}\frac{\Gamma\left(\frac{ix}{Q}\right)^2}{\Gamma\left(1-\frac{ix}{Q}+|N| + i N \frac{\Theta}{Q}\right)^2}H_N(x+i\,|N|\,\frac{Q}{2}-N \frac{\Theta}{2})\;\;\; .
\end{eqnarray}
Here we have defined $H_N(x)$ by
\begin{eqnarray}
H_N(x)&=&\prod_{m\neq n+N}\frac{((m+1 )b+(n+1)/b-ix)^2}{(mb+n/b+ix)^2}\,.
\end{eqnarray}

Note that the reason for this splitting into ratio of gamma functions and $H_N(x)$ arises quite naturally by using equations (\ref{eq:Theta})-(\ref{eq:action})
to rewrite the one-loop determinant (\ref{eq:sbproduct})
\begin{align}
s_b(x+i Q/2) &= \prod\limits_{B\in\mathbb{Z}} \prod\limits_{k=0}^\infty \frac{|B| Q +i B\, \Theta +2(k+1) Q-2ix}{|B| Q+i B\,\Theta+2k Q+2ix}\\
&=\prod\limits_{B\in\mathbb{Z}} \frac{\Gamma\left( \frac{i x}{Q} + \frac{|B|}{2}+i \frac{B\,\Theta}{2Q}\right)}{\Gamma\left(1- \frac{i x}{Q} + \frac{|B|}{2}+i \frac{B\,\Theta}{2Q}\right)}\,,
\end{align}
where we defined $B= m-n$ and $k=\mbox{min}(m,n)$.
In the given $N^{th}$ topological sector we factorise out the ratio of gamma functions coming from the $B=N$ term which will be the only singular factor along the corresponding contour of integration; everything else is collected in these auxiliary functions $H_N(x)$. Once more if we were to set $H_N(x)\to 1$ we would obtain precisely the contribution coming from the topological sector with magnetic flux $B=N$ and $\theta$ angle directly related to our $\Theta$ for the two-dimensional supersymmetric $\mathbb{CP}^1$ model discussed in \cite{Dorigoni:2017smz}.

Here as well we can regulate the function $H_N(x)$ using q-Pochhammers, but as it is entire along the contour and in its interior we will not need its precise form. When we can evaluate these integrals we get an expansion of the form
\begin{eqnarray}\label{eq:TSN}
\zeta_N(\xi,b)=\sum\limits_{k=0}^\infty e^{-2\pi \xi k Q}\zeta_{N,k}(\xi,b) \;\;\; ,
\end{eqnarray}
precisely as expected from our resurgence triangle discussion for equation (\ref{eq:generalTS}).
As seen for the topologically trivial sector, when we write $\zeta_{N,k}(\xi,b)$ as a perturbative series in $\xi$ and we find that it truncates after finitely many orders. In the present case of two chirals the truncation happens precisely after two orders, so we will need to deform the theory before we are able to apply the resurgence framework to reconstruct non-perturbative information from perturbative data.

\subsection{Cheshire Cat deformation}

To re-introduce the (general) asymptotic nature of every perturbative expansion we now want to add a Cheshire cat deformation to the theory. Following \cite{Dorigoni:2017smz} we have two options to consider. One possibility is to analytically deform the integrand of the localised partition function to mimic a non-supersymmetric unbalance between the number of bosonic and fermionic degrees of freedom. To do this we note that the matter one-loop determinant for the chiral theory $(N_c,0)$ can easily be written as
\begin{eqnarray}
Z_{matter}=\left(\frac{\mathrm{det}O_\psi }{\mathrm{det}O_\phi}\right)^{N_c}\;\;\; .
\end{eqnarray}
Thus this supersymmetry breaking deformation would look something like
\begin{eqnarray}
{\tilde{Z}}_{matter} =  \frac{ \left( \mbox{det} \mathcal{O}_\psi\right)^{N_f}}{\left(\mbox{det} \mathcal{O}_\phi\right)^{N_b}}=Z_{matter} \,\left(\mbox{det} \mathcal{O}_\phi\right)^{\Delta}\; ,
\end{eqnarray}
where we have set $N_b=N_c-\Delta$ and $N_f=N_c$. To proceed we would need to keep and regulate the full one-loop determinant written as an infinite product over eigenvalues with degeneracies, which can be found in \cite{Hama:2011ea}, \cite{Fujitsuka:2013fga}, without all the cancellations between pairing of bosonic and fermionic modes that take place when $\Delta=0$ producing the simpler expression (\ref{eq:sbproduct}).

The second option, which turns out to be nicer, is to deform the number of chiral multiplets to be non-integer, $N_c\rightarrow N_c+\Delta$. Everything we will discus in this paper works perfectly fine in both cases, but the expressions are much shorter for this latter deformation, and just as illuminating. In this case we have
\begin{eqnarray}
{\tilde{Z}}_{matter} = \left( \frac{ \mbox{det} \mathcal{O}_\psi}{\mbox{det} \mathcal{O}_\phi}\right)^{N_c+\Delta}\;\;\; .
\end{eqnarray}
Because both the fermionic and bosonic determinants are raised to the same power we still have the same cancellations between the determinants, and so we can stick with the one-loop determinant expressions we already have.

In this Section we will focus only on this second type of Cheshire Cat deformation where we analytically continue in the number of chiral fields to non-integer values. In \cite{Dorigoni:2017smz} we have already shown in a $2$-d context how the introduction of an unbalance between bosons and fermions, effectively breaking supersymmetry, produces very similar results.

However a striking point we want to stress is how almost any conceivable deformation of the theory will immediately make the perturbative expansions asymptotic, allowing us to use the full machinery of resurgent analysis. In many cases, supersymmetrically localised theories are effectively sitting at very special points in the space of ``physical functions'' where miraculous cancellations hide the resurgence structure. Whenever a Cheshire Cat deformation is re-instated we can instantly see reappearing the complete resurgent body, and when taking the vanishing limit of this deformation only its grin will remain.

For simplicity we will now work in the topologically trivial sector, but everything follows through in the other sectors in exactly the same manner, one has just to replace $H_0(x)$ by $H_N(x+iNb)$ or $H_N(x-iN/b)$ and the ratio of gamma according to (\ref{eq:TopSecIntegral}).
Now applying the deformation instead of equation (\ref{eq:zerosec}) we get
\begin{align}
\Tilde{\zeta}_0(\xi,b,\Delta)&=\int\limits_{\Gamma_0} dx \,e^{2\pi i x \xi}\frac{\Gamma\left(\frac{ix}{Q}\right)^{\Delta+2}}{\Gamma\left(1-\frac{ix}{Q}\right)^{\Delta+2}}H_0(x) \\
&\nonumber=\int\limits_{\Gamma_0} dx\, e^{2\pi i x \xi} H_0(x) e^{(\Delta+2)\left[\log\Gamma\left(\frac{ix}{Q}\right) -\log\Gamma\left(1-\frac{ix}{Q}\right)\right]}\,. 
\end{align}
Note that in principle the deformation would also alter the function $H_0(x)\to H_0(x)^{1+\Delta/2}$. However this turns out to be superfluous since the deformation of $H_0(x)$ will not add anything new and to recover the resurgence structure it will be sufficient to just deform the ratio of gamma functions. The only change we want to point out is that both the poles and the zeroes of $H_0(x)$ will become branching points for $H_0(x)^{1+\Delta/2}$.

Now the contour $\Gamma_0$ comes down from $+i\infty-\epsilon$, circles the origin and goes back up to $+i\infty+\epsilon$. We make the change of variables $x\to i x$ so the integral is now along the positive real axis and we note that the function $\log\Gamma(-\frac{x}{Q})$ has a branch cut precisely on the contour of integration so we obtain the integral along the real axis of its discontinuity
\begin{eqnarray}
\Tilde{\zeta}_0(\xi,b,\Delta)=i\int\limits_0^\infty&& dx\,e^{-2\pi\xi x}H_0(ix)e^{-(\Delta+2)\log\Gamma\left(1+\frac{x}{Q}\right)} \\
&&\left(e^{(\Delta+2)\log\Gamma\left(-\frac{x}{Q}+i\epsilon\right)}-e^{(\Delta+2)\log\Gamma\left(-\frac{x}{Q}-i\epsilon\right)}\right)\;\;\; .
\end{eqnarray}
We can now use the discontinuity formula,
\begin{eqnarray}
\log \Gamma(-x+i\epsilon)-\log \Gamma(-x-i\epsilon)=-2\pi i\left(\lfloor x\rfloor+1\right)\,,
\end{eqnarray}
where $\lfloor x\rfloor$ denotes the floor of $x$,
to write $\Tilde{\zeta}_0(\xi,b,\Delta)$ in the form
\begin{eqnarray}\label{deformed0topo}
\Tilde{\zeta}_0(\xi,b,\Delta)=i\int\limits_0^\infty &&dx\;e^{-2\pi\xi x}H_0(ix)e^{-(\Delta+2)\log\Gamma\left(1+\frac{x}{Q}\right)+(\Delta+2)\log\Gamma\left(-\frac{x}{Q}\pm i\epsilon\right)} \\
&&e^{\pm \pi i(\Delta+2)\left(\lfloor\frac{x}{Q}\rfloor+1\right)}\left(e^{- \pi i(\Delta+2)\left(\lfloor\frac{x}{Q}\rfloor+1\right)}-e^{+ \pi i(\Delta+2)\left(\lfloor\frac{x}{Q}\rfloor+1\right)}\right)\;\;\; . \nonumber
\end{eqnarray}
Next to make manifest the transseries nature of this integral we rewrite the domain of integration as
\begin{equation*}
\int\limits_0^\infty dx\; f(x)=\sum\limits_{k=0}^\infty\,\int\limits_{kQ}^{(k+1)Q} dx\;f(x) \,,
\end{equation*}
evaluate the floor function on each interval and then use the identity
\begin{equation}
\sum\limits_{k=0}^\infty\,\int\limits_{kQ}^{(k+1)Q} dx\;f(x)= \sum\limits_{k=0}^\infty\left(\,\int\limits_{kQ}^{\infty} dx\;f(x)-\int\limits_{(k+1)Q}^{\infty} dx\;f(x)\right) \;.
\end{equation}
Finally we change variables to make all the integrals start from the origin.
In this way we can write
\begin{eqnarray}\label{eq:TSdeformed}
\Tilde{\zeta}_0(\xi,b,\Delta)=\sum\limits_{k=0}^\infty e^{-2\pi \xi k Q}\tilde{\zeta}_{0,k}(\xi,b,\Delta)\;\;\; .
\end{eqnarray}
For the moment we specialise to $\Tilde{\zeta}_{0,0}(\xi,b,\Delta)$ which takes the form 
\begin{eqnarray}
\Tilde{\zeta}_{0,0}(\xi,b,\Delta)=2\sin(\pi\Delta)e^{\pm i \pi \Delta} \int\limits_0^\infty dx \,e^{-2\pi\xi x}H_0(ix)e^{-(\Delta+2)\left[ \log\Gamma\left(1+\frac{x}{Q}\right)-\log\Gamma\left(-\frac{x}{Q}\pm i\epsilon\right)\right]}\, , \nonumber
\end{eqnarray}
and by making use of the shift formula
\begin{eqnarray}\label{shiftformula1}
\log\Gamma(-x\pm i\epsilon)=\log\Gamma(1-x\pm i\epsilon)-\log(x)\mp i\pi\;\; ,
\end{eqnarray}
we obtain
\begin{eqnarray}
\Tilde{\zeta}_{0,0}(\xi,b,\Delta)&=&2\sin(\pi\Delta)\int\limits_0^\infty dx\,e^{-2\pi\xi x}H_0(ix)\left(\frac{x}{Q}\right)^{-(\Delta+2)} e^{-(\Delta+2)\left[\log\Gamma\left(1+\frac{x}{Q}\right)-\log\Gamma\left(1-\frac{x}{Q}\pm i \epsilon\right)\right]} \nonumber \\
&=&\label{eq:LatLapPhi0}\int\limits_0^\infty dx\,e^{-2\pi\xi x}x^{-(\Delta+2)} \Phi^{(0)}_0(x\mp i \epsilon)=  \tilde{\mathcal{S}}_{\mp}\left[\Phi_0^{(0)}\right](\xi,b,\Delta)\,.
\end{eqnarray}
In the last line we introduced the modified lateral Laplace transform whose explicit definition is given by
\begin{align}
\tilde{\mathcal{S}}_{\mp}\left[\Phi\right](\xi) &\notag= \lim_{\epsilon \to 0^+}\int\limits_0^{\infty\mp i \epsilon} dx\,e^{-2\pi\xi x}x^{-(\Delta+2)} \Phi(x)\\
&\label{eq:LapTr}= \lim_{\epsilon \to 0^+}\int\limits_0^{\infty} dx\,e^{-2\pi\xi x}x^{-(\Delta+2)} \Phi(x\mp i\epsilon)\,.
\end{align}
The Borel transform of the purely perturbative part in the topologically trivial sector $\Phi^{(0)}_0(x)$ can be read off from (\ref{eq:LatLapPhi0}) 
\begin{equation}\label{eq:Phi0}
\Phi^{(0)}_0(x) = 2\sin(\pi\Delta) \,H_0(ix)\,Q^{\Delta+2} \,e^{-(\Delta+2)\left[\log\Gamma\left(1+\frac{x}{Q}\right)-\log\Gamma\left(1-\frac{x}{Q}\right)\right]}\,.
\end{equation} 
Importantly this has finite radius of convergence around the origin and can be expanded as a power series in $x$
\begin{eqnarray}
\Phi^{(0)}_0(x)=\sin(\pi\Delta)\sum\limits_{m=0}^\infty c_{0,0,m}(b,\Delta)x^m\; .
\end{eqnarray}
After commuting this series with the Laplace integral we finally obtain
\begin{eqnarray}\label{eq:PertAsy}
\tilde{\zeta}_{0,0}(\xi,b,\Delta)=\sin(\pi\Delta)(2\pi\xi)^{\Delta+2}\sum\limits_{m=0}^\infty \frac{c_{0,0,m}(b,\Delta)}{(2\pi\xi)^{m+1}}\Gamma(m-1-\Delta) \;\;\; .
\end{eqnarray}
It is simple to note that for generic, i.e. non-integer $\Delta$ this series is asymptotic. Precisely as anticipated after having performed this Cheshire Cat deformation the perturbative expansion is not truncating any longer and we are left with a factorially growing asymptotic series. 
Furthermore when we take the limit $\Delta\rightarrow 0$ we have $\sin(\pi\Delta)\rightarrow 0$ but simultaneously the $\Gamma(m-1-\Delta)$ develops poles for $m=0,1$, hence in this limit we reproduce exactly the undeformed result $\tilde{\zeta}_{0,0}(\xi,b,\Delta)\stackrel{\Delta\to 0}{\longrightarrow}\zeta_{0,0}(\xi,b)$ of equation (\ref{undeformedcoefficients}).

We can also find the general expression for all $\tilde{\zeta}_{0,k}(\xi,b,\Delta)$. Starting from equation (\ref{deformed0topo}), rewriting the integral as we did before, and using the shift formula (\ref{shiftformula1}) we find
\begin{eqnarray}
\Tilde{\zeta}_{0,k}(\xi,b,\Delta)&=&\notag2\sin(\pi\Delta) e^{\pm i \pi k \Delta } \int\limits_0^\infty  dx\,e^{-2\pi\xi x}H_0(ix+ikQ )  \prod\limits_{n=0}^k \left(\frac{x}{Q}+n \right)^{-(\Delta+2)}  \\
&&\nonumber \phantom{2\sin(\pi\Delta) e^{\pm i \pi k \Delta } \int\limits_0^\infty  dx}e^{-(\Delta+2)\left[ \log\Gamma\left(1+k+\frac{x}{Q}\right)-\log\Gamma\left(1-\frac{x}{Q}\pm i\epsilon\right)\right]} \nonumber \\
=&&e^{\pm i \pi k \Delta }\int\limits_0^\infty dx\;e^{-2\pi\xi x} x^{-(2+\Delta)}\Phi_0^{(k)}(x\mp i\epsilon) \nonumber \\
=&&\label{eq:NPintegral}e^{\pm i \pi k \Delta }\tilde{\mathcal{S}}_{\mp}\left[\Phi_0^{(k)}\right](\xi,b,\Delta)\;\;\; ,
\end{eqnarray}
where once more we used the modified lateral Laplace transform (\ref{eq:LapTr}) to integrate the Borel transform $\Phi_0^{(k)}(x)$ of the $k^{th}$ vortex-anti-vortex non-perturbative sector that reads
\begin{align}
\Phi_0^{(k)}(x) =&\notag 2\sin(\pi\Delta) \,H_0(ix+ikQ )\,Q^{\Delta+2}  \prod\limits_{n=1}^k \left(\frac{x}{Q}+n \right)^{-(\Delta+2)} \\
&\label{eq:Phik}e^{-(\Delta+2)\left[ \log\Gamma\left(1+k+\frac{x}{Q}\right)-\log\Gamma\left(1-\frac{x}{Q}\right)\right]}\,,
\end{align}
reducing to (\ref{eq:Phi0}) for $k=0$.

Similarly to the purely perturbative series also in the non-perturbative sectors one can expand the Borel transform as a convergent power series at the origin $x=0$ and commute the sum with the integral to obtain an asymptotic, factorially growing power series for generic $\Delta$. Taking the limit $\Delta\to0$ reproduces precisely the truncating perturbative series (\ref{undeformedcoefficients}) of the undeformed case.

Putting everything together we arrive at the complete transseries expression for (\ref{eq:TSdeformed})
\begin{equation}
\Tilde{\zeta}_0(\xi,b,\Delta)=\sum\limits_{k=0}^\infty e^{-2\pi \xi k Q}  e^{\pm i \pi k \Delta } \tilde{\mathcal{S}}_{\mp}\left[\Phi_0^{(k)}\right](\xi,b,\Delta)\,,\label{eq:1ParamTS}
\end{equation}
where the factor $e^{\pm i \pi k \Delta }$ is called the transseries parameter.
Note that a similar analysis can be carried out in each topological sector.

The expression (\ref{eq:1ParamTS}) for the full, perturbative and non-perturbative, set of contributions to the topologically trivial sector tells us that we are working with what is called a one parameter transseries.
One might think that according to our choice of sign $e^{ i \pi k \Delta } \tilde{\mathcal{S}}_{-}$ or $e^{ -i \pi k \Delta } \tilde{\mathcal{S}}_{+}$ we would find two different results for real and positive $\xi$; however as was shown in full details in \cite{Aniceto:2013fka} for the most general one parameter the jump in this transseries parameter is precisely needed to cancel the ambiguity in the resummation $(\tilde{\mathcal{S}}_{+}-\tilde{\mathcal{S}}_{-})[\Phi_0^{(k)}]$, also called Stokes automorphism.

Our transseries (\ref{eq:1ParamTS}) is completely real and unambiguous for real and positive $\xi$: one can use the analysis\footnote{Note that in the present case the function $H_0$ does not really play any role and it is just carried along the way. The one-parameter nature of the transseries under consideration comes from the particular combination of $\log\Gamma$ functions.} of Section 6 of \cite{Dorigoni:2017smz} or the more general expressions in \cite{Aniceto:2013fka} to show that the would-be ambiguity cancels order by order in the vortex-anti-vortex counting parameter $e^{-2\pi \xi k Q}$.

The imaginary part of the transseries parameter $\mbox{Im} e^{ \pm i \pi k \Delta } = \pm \sin (\pi k \Delta)$ is exactly (anti-)correlated with the discontinuity $(\tilde{\mathcal{S}}_{+}-\tilde{\mathcal{S}}_{-})[\Phi_0^{(k)}]$. Hence (\ref{eq:1ParamTS}) is the real solution corresponding to what is called median resummation (see \cite{Delabaere1999} and the general discussion in \cite{Aniceto:2013fka}).
Using the results of Section 6 of \cite{Dorigoni:2017smz} we can also rewrite (\ref{eq:1ParamTS}) in the manifestly real and unambiguous form
\begin{eqnarray}
\Tilde{\zeta}_0(\xi,b,\Delta)&=& \sum\limits_{k=0}^\infty e^{-2\pi \xi k Q} \cos^k(\pi\Delta) \,\tilde{\mathcal{S}}_0\left[\mathrm{Re}\left(\Phi^{(k)}_0\right)\right](\xi,b,\Delta)\nonumber \\
&=&\tilde{\mathcal{S}}_0\left[\mathrm{Re}\left(\Phi^{(0)}_0\right)\right](\xi,b,\Delta)+e^{-2\pi \xi Q}\cos(\pi\Delta)\,\tilde{\mathcal{S}}_0\left[\mathrm{Re}\left(\Phi^{(1)}_0\right)\right](\xi,b,\Delta)\nonumber \\
&&+e^{-4\pi \xi Q}\cos^2(\pi\Delta)\,\tilde{\mathcal{S}}_0\left[\mathrm{Re}\left(\Phi^{(2)}_0\right)\right](\xi,b,\Delta)+O\left(e^{-6\pi \xi Q}\right)\,,
\end{eqnarray}
where $\tilde{\mathcal{S}}_0$ denotes the modified Laplace transform (\ref{eq:LapTr}) where the integration contour is the positive real axis which we can do now given the fact that $\mathrm{Re}\left(\Phi^{(k)}_0\right)(x)$ is completely regular for $x>0$.

As already stressed if we were to expand each Laplace integral as a power series we would obtain a factorially divergent perturbative expansion in $1/\xi$ in each non-perturbative sector, however when we take the limit $\Delta\to 0$ all of these will truncate to finitely many perturbative coefficients thus reproducing (\ref{undeformedcoefficients}). We will now show that having made the body of the Cheshire Cat visible by considering generic $\Delta$ will allow us to reconstruct the non-perturbative sectors from the asymptotic perturbative one and vice-versa.

\subsection{Non-perturbative data from perturbation theory}

What we would like to do now is using the resurgence machinery to reconstruct the deformed non-perturbative sectors (\ref{eq:NPintegral}) and eventually the undeformed contributions (\ref{undeformedcoefficients}) from the deformed resummed perturbative data (\ref{eq:LatLapPhi0}) or equivalently from the deformed asymptotic perturbative series (\ref{eq:PertAsy}). 

A standard method is to start from the perturbative asymptotic power series (\ref{eq:PertAsy}) and resum it by performing a directional Laplace integral of its Borel transform (\ref{eq:Phi0}) 
\begin{eqnarray}\label{eq:directional}
(2\pi \xi)^{-(\Delta+2)}\Tilde{\zeta}_{0,0}(\xi,b,\Delta)&=& \int\limits_0^{\infty }\frac{dy}{2\pi \xi} e^{-y} y^{-(\Delta+2)} \Phi^{(0)}_0\left(\frac{y}{2\pi \xi}\right)  \\
&=&\nonumber \int\limits_0^{\infty e^{-i\theta} } dx\, e^{-2\pi \xi x} ( 2\pi \xi x)^{-(\Delta +2)} \Phi^{(0)}_0(x)=\mathcal{S}_\theta [\Phi^{(0)}_0](\xi,b,\Delta)\,,
\end{eqnarray}
where $\theta = \arg \xi$ and $\mathcal{S}_\theta $ denotes the modified directional Laplace transform, similar to equation (\ref{eq:LapTr}) (in here we added an additional factor $(2\pi x)^{-(\Delta+2)}$ for convenience).

The above equation does define a function on the complex variable $\xi$ by anti-correlating its argument with the direction of the Laplace transform. This function is defined everywhere on the complex $\xi$ plane save some cuts where there is a discontinuity in the directed Laplace transform because of singularities of the integrand, i.e. the Stokes directions of the Borel transform. 

A well known dispersion like argument \cite{Bender:1990pd,Collins:1977dw} applied to the function just constructed from the purely perturbative data, i.e. $\Tilde{\zeta}_{0,0}(\xi,b,\Delta)$, would generically allow us to relate the asymptotic form of the perturbative coefficient (\ref{eq:PertAsy}) to the discontinuities of this function, which in turn directly relates to all the non-perturbative contributions (\ref{eq:NPintegral}) (and their associated perturbative expansions) coming from the tower of $k$ vortex-anti-vortex configurations in the same topological sector.

In the present case however we cannot straightforwardly use this standard method because of presence of the function $H_0(ix)$ within the Borel transform (\ref{eq:Phi0}). This function has poles, or alternatively its Cheshire Cat deformation, $H_0(ix)^{1+\Delta/2}$, has branch cuts going out horizontally to infinity in the positive real direction starting at $ x=m b$ and $x=m/b$ for $m\in\mathbb{N}^*$ as one can read from the denominator of (\ref{eq:H0}).
For this reason in equation (\ref{eq:directional}) there are no straight rays emanating from the origin $x=0$ in a direction $\theta$ with $-\arg  b\leq \theta \leq \arg b$ without intersecting any of the singular directions.

\begin{figure}[ht]
\centering
\includegraphics[width=0.6\textwidth]{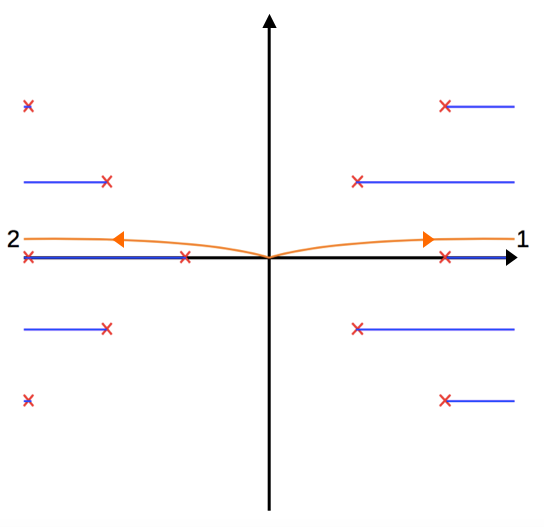}
\caption{Contours of integration for the directed Laplace transformations.}
\label{LargeOrder1}
\end{figure}

This suggests that we just need to find a different way from (\ref{eq:directional}) to define a function of the complex $\xi$ variable with countably many branch cuts.
One such way is as follows. We define this function by gluing analytic functions defined in different wedges of the complex $\xi$ plane. First we consider the directional Laplace contour along the first integration contour shown in Figure \ref{LargeOrder1}. This defines a function of $\xi$ analytic for $-\pi/2<\arg \xi<\pi/2$. Likewise we use the second path shown in Figure \ref{LargeOrder1} to define a function of $\xi$ in the wedge $\pi/2<\arg \xi<\pi$ union with $-\pi<\arg \xi<-\pi/2$. 
The function thus obtained will have branch cuts along the directions $\arg \xi = \pm \pi/2$ and its discontinuities will be related to the discontinuity of the Borel transform along the directions $\arg x =0$ and $\arg x= \pi$ which in turn are related to all the $k$ vortex-anti-vortex non-perturbative sectors, but also infinitely many other discontinuities associated with $H_0(ix)^{1+\Delta/2}$ with starting points either the poles or the zeroes of (\ref{eq:H0}).

This is somewhat unexpected from the resurgence point of view since these additional branch cuts are associated with \textit{different} topological sectors from the one we were focusing on! 
In resurgence theory when we work in a given topological sector, say for example the trivial one, we complexify the coupling constant to understand the analytic properties of the resummed perturbative series and from this reconstruct the non-perturbative contributions in the \textit{same} topological sector. Said in other words the imaginary part of the complexified coupling constant has nothing to do with the topological angle. Hence on resurgent ground we generically expect the Borel transform of the purely perturbative data to know ``everything'' about non-perturbative saddles in the same topological sector and ``nothing'' about different topological sectors.
This is of course if no other structure is present as we will discuss in the next Session.

The present case is entirely different and the reason behind it lies in the unusual appearance of the hidden topological angle and the path-integral decomposition in topological sectors (\ref{toposplit}).
The imaginary part of the action in the $N^{th}$ topological sector is given by $\mbox{Im} S\propto \xi \Theta N \propto (b-1/b) \xi$.
Now it is clear that what we just said is not true anymore; if we keep fixed $\Theta=  -i(b-1/b)$, complexify the coupling constant $\xi$ and vary its imaginary part we will inevitably vary the theta angle, i.e. the imaginary part of the action of each topological sector. Hence in the case at hand we have some additional structure (see more in the next Section), making so that the Borel transform of the purely perturbative data knows also of different topological sectors.

We found however two different methods that can be applied to these standard dispersion arguments to disentangle from the Borel transform the branches coming from the same topological sector and the ones coming from other sectors. As a proof of principle we will now present both but will not dwell too much on the consequences.

A first possibility is to impose that, as a genuine theta angle would do, indeed $\mbox{Im} S =  \pi \xi \Theta N \sim (b-1/b) \xi$ is independent from the complexification of $\xi$. If we assume the double scaling limit $\xi\to \infty$ and simultaneously $b=e^{i \vartheta /\xi}$ we have that $\mbox{Im} S \propto \vartheta$ is independent from $\xi$. In this regime when we complexify $\xi$ we have that $b$ is not of unit modulus anymore; however the background geometry discussed in Section \ref{physicalinterpritation} still makes sense.
The price to pay is that now the weak coupling expansion $\xi \to \infty$ of (\ref{eq:LatLapPhi0}) will not be as straightforward as when we computed the factorially growing perturbative series (\ref{eq:PertAsy}) since in this double scaling limit $b$ is no longer an independent parameter and the Borel transform does depend from the coupling through $b$.

An alternative method is to define something similar to (\ref{eq:directional}) but not holomorphic:
\begin{eqnarray}\label{resummation2}
(2\pi \xi)^{-(\Delta+2)}\Tilde{\zeta}_{0,0}(\xi,b,\Delta)&=&
2\sin(\pi\Delta)\int\limits_0^{\infty } \frac{dy}{2\pi \xi}\, e^{-y} \,\left(\frac{y}{Q}\right)^{-(\Delta+2)}\, H_0\left(\frac{i y}{|\xi| }\right)\nonumber \\
&&\nonumber\phantom{2\sin(\pi\Delta)\int}e^{-(\Delta+2)\left( \log\Gamma\left(1+\frac{y}{2\pi \xi Q}\right)-\log\Gamma\left(1-\frac{y}{2\pi \xi Q}\right)\right)}\\
&&=2\sin(\pi\Delta)\int\limits_0^{\infty e^{-i \theta}}  dx\, e^{-2\pi\xi x}\,\left(\frac{2\pi \xi x}{Q}\right)^{-(\Delta+2)}H_0(ix e^{+ i\theta})\nonumber \\
&&\phantom{2\sin(\pi\Delta)\int}e^{-(\Delta+2)\left( \log\Gamma\left(1+\frac{x}{Q}\right)-\log\Gamma\left(1-\frac{x}{Q}\right)\right)},
\end{eqnarray}
where again $\theta = \arg\, \xi$ and we anti-correlate the direction of the Laplace transform with the argument of the complexified coupling constant.
The difference is that as we rotate the argument of $\xi$ we simultaneously rotate the branches of the function $H_0(ix)$ so that they never cross our contour of integration, or equivalently in the $y$ variable as we rotate the argument of $\xi$ the only branches crossing the contour of integration are the ones coming from the $\log \Gamma$ functions and not from $H_0( i y)$.
Hence as a function of $\xi$ we only have two discontinuities now, one across the $\arg \xi = 0$ direction which will persist the $\Delta \to 0$ limit and one across the $\arg \xi = \pi$ direction which will disappear in the $\Delta \to 0$ limit.

With this definition we still have exactly the same perturbative asymptotic series (\ref{eq:PertAsy}) since for $\xi >0$ we trivially have that $\xi = | \xi|$. However when performing the Borel transform we treat differently terms coming belonging to the same topological sector from terms belonging to others in what effectively seems like an extremely ad-hoc prescription.

As mentioned before these discontinuities will be related to non-perturbative contributions and with this, once again very a posteriori, prescription we can isolate only the non-perturbative saddles in the same topological sector. It would be nice to provide some numerical examples of large order relations similar to \cite{Bender:1990pd,Collins:1977dw}, but unfortunately this turns out to be quite non-trivial. The main issue we have with running some numerics lies in evaluating the function $H_0(ix)$, at $x=0,Q,2Q,...$ and so on. Using the results outlined in Appendix \ref{doublesine} this should be a doable task but we have decided to be content with the analytic results derived and defer the numerics to future works.

\section{Relation between different topological sectors}\label{dunneunsalsection}

\begin{figure}[ht]
\centering
\includegraphics[width=0.8\textwidth]{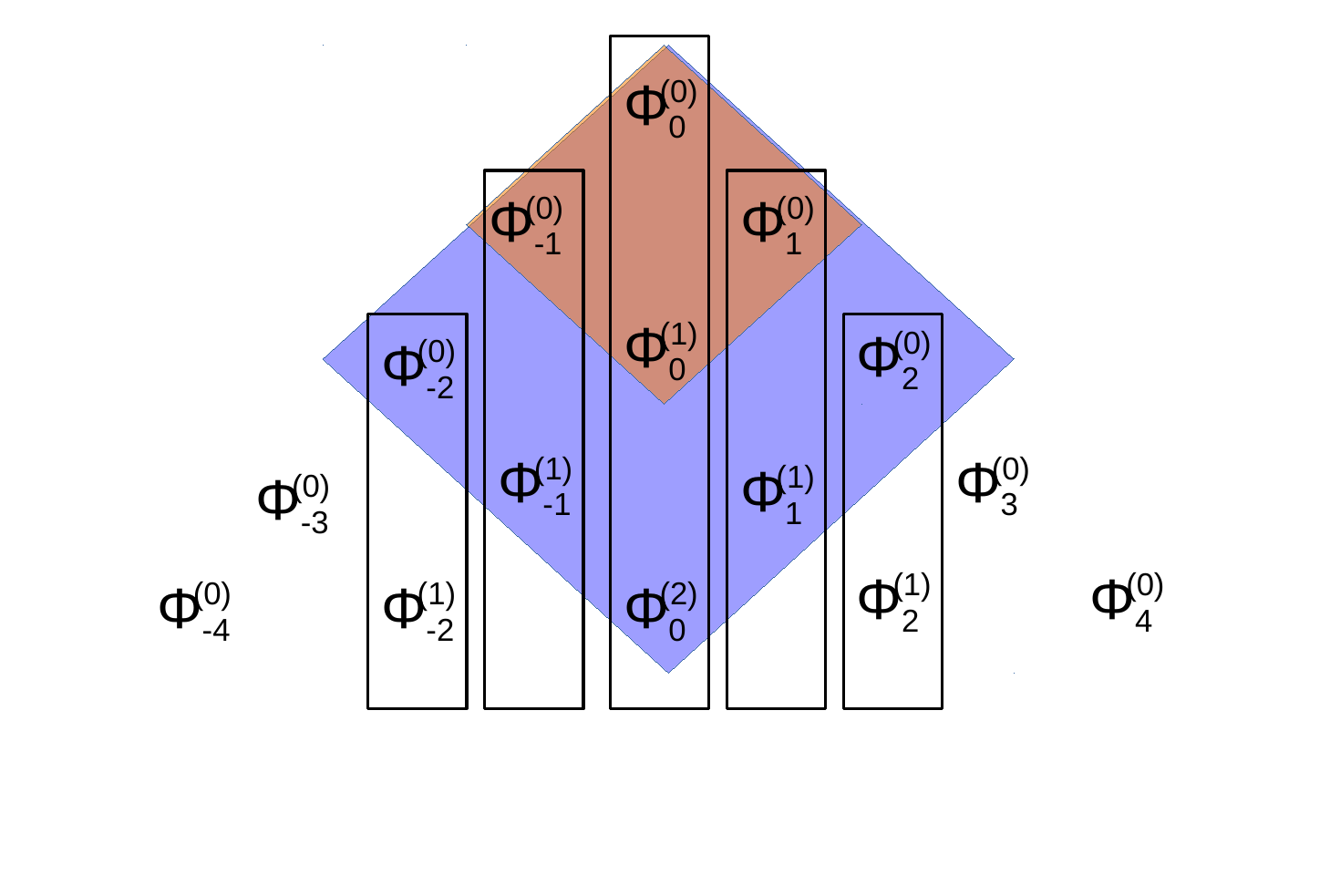}
\caption{Resurgence theory allows us to move vertically along each rectangle. From just one of the contributions in a topological sector, i.e. $\Phi^{(k)}_B$ we can get all the $\Phi^{(k')}_B$ with $k'\neq k$.}
\label{DunneUnsal}
\end{figure}

So far we have understood that the localised partition function can be written as a transseries (\ref{eq:TopoTS}) over different topological sectors for which the imaginary part of the squashing parameter plays the role of a hidden topological angle. Each topological sector can be furthermore written as a transseries (\ref{eq:TSundef}) capturing the perturbative series in the given topological sector, plus the infinitely many non-perturbative contributions coming from vortex-anti-vortex configurations on top of it. Upon Cheshire Cat deformation (\ref{eq:TSdeformed}) from a given perturbative series we can reconstruct every element in the same topological sector, i.e. from one element of the resurgent triangle of Figure \ref{DunneUnsal} we can reconstruct all the other elements in the same column. In this Section we wish to discuss the relation between the topological sectors and additional structures allowing us to move ``horizontally'' in the resurgence triangle.

As stressed in the previous Section the theta angle can be seen as introducing a grading in the partition function (\ref{eq:TopoTS}), a sort of Fourier modes decomposition. Once we work topological sector by topological sector we complexify the coupling constant and use resurgence theory to understand its analytic properties, but the theta angle and the imaginary part of the coupling must not be confused with one another. To be able to move between different topological sectors we need some additional structure that somehow links the theta angle to the complexified coupling constant.

In many supersymmetric QFTs we indeed have this type of additional structure which allows us to use the data contained in the transseries in the trivial topological sector, e.g. for the present case (\ref{eq:TSundef}), to calculate the data in different topological sectors.

For example in $2$-dimensional $\mathcal{N}=(2,2)$ supersymmetric field theories we have shown \cite{Dorigoni:2017smz} how the $tt^\star$ structure of Cecotti and Vafa \cite{Cecotti:1991me} is modified but still imposes that the partition function must satisfy a differential equation in the holomorphic coupling $\tau\sim \xi +i\theta$. 
This is precisely the extra structure needed.
With resurgence theory we complexify $\xi$ to reconstruct for example the topologically trivial sector from perturbation theory and then use the $tt^\star$ differential equation to obtain the data for sectors with non-trivial $\theta$ dependence; i.e. the data for the instanton and anti-instanton sectors are intimately tied up with the instanton-anti-instanton contributions, and so on. From the $n$-instanton-$n$-anti-instanton contribution we can use this additional structure to calculate the contributions from different sectors in the resurgence triangle moving ``horizontally'' across the resurgence triangle as shown in Figure \ref{DunneUnsal}.

Schematically, if we were to reconstruct from the perturbative data $\Phi_0^{(0)}$ say the first instanton-anti-instanton contribution $\Phi_0^{(1)}$ we would then be able to retrieve all the data in the red square of Figure \ref{DunneUnsal}. Similarly once we reconstruct the 2-instanton-2-anti-instanton sector $\Phi_0^{(2)}$ out of perturbation theory we would have access to the entire blue square of Figure \ref{DunneUnsal}. 

This is very reminiscent of the Dunne-\"Unsal relation in quantum mechanics \cite{Dunne:2014bca,Dunne:2013ada,Gahramanov:2015yxk}. There the same could be achieved; the data in the instanton-anti-instanton, or even just in the purely perturbative series, can be related to the data in the instanton sector. In that case the relationship was derived using boundary conditions on the non-perturbative effects.

In the present $3$-d $\mathcal{N}=2$ case we have a very similar story. Here we have what is usually called vortex-anti-vortex factorisation discussed in \cite{Pasquetti:2011fj}. 
Reminiscent of the $2$-d case \cite{Benini:2012ui,Doroud:2012xw}, these three dimensional theories, say for example with $2N$ chirals, have a partition function that factorises schematically as
\begin{eqnarray}\label{factorisation}
Z^{(2N,0)}_{S_b^3}(\xi)=\sum\limits_{i=1}^N Z_{cl}^{(i)}\times \left(Z_{1-loop}^{(i)}Z_{V}^{(i)}\right)\times \left(\Bar{Z}_{1-loop}^{(i)}\Bar{Z}_{V}^{(i)}\right)\;\;\; .
\end{eqnarray}
Here $Z_{cl}^{(i)} =\exp( -i \pi \xi_{\tiny{\mbox{eff}}}\,\mu_i)$ is the classical part of the action, with $\mu_i$ axial mass for the $i^{th}$ chiral and $ \xi_{\tiny{\mbox{eff}}}$ the effective FI parameter, while $Z_{V}^{(i)}$ and its complex conjugate are the abelian vortex and anti-vortex partition functions with $2N$ chirals, dressed by $Z_{1-loop}^{(i)}$ and its conjugate.

As discussed in \cite{Pasquetti:2011fj} the vortex partition function $Z_V$ can be better understood in the degenerate $b\to 0$ limit where the background geometry becomes $\mathbb{R}^2\times S^1$ and the partition function counts finite-energy configurations on $\mathbb{R}^2$, i.e. vortices. Similarly in the $1/b\to 0$ limit the squashed sphere degenerates to a different $\mathbb{R}^2\times S^1$ and the partition function $\Bar{Z}_{V}$ counts anti-vortices.
 
From this factorised form it is now not surprising that the transseries in different topological sectors are related to one another.
Hence we have the following method to obtain all the non-perturbative data in all the topological sectors from the perturbative data alone. First we deform the theory with some Cheshire Cat deformation as to re-introduce all the asymptotic tails in the various perturbative expansions. Next we use usual resurgence methods on the deformed theory to calculate all the non-perturbative vortex-anti-vortex contributions in the trivial topological sector. Then we send the deformation back to zero, retrieving all the non-perturbative data in this sector for the undeformed theory. Finally we use the factorisation formula (\ref{factorisation}) to compute the data for all the other topological sectors from the non-perturbative data in the trivial topological sector.

It is interesting to push this idea to higher dimensions. In fact it was already noted in \cite{Pasquetti:2011fj} that this factorised form for the partition function (\ref{factorisation}) is very reminiscent of the Nekrasov structure in $4$-d. 
If we focus for example to $4$-d $\mathcal{N}=2$ theories on $S^4$ we have Pestun's celebrated partition function \cite{Pestun:2007rz}
\begin{eqnarray}
Z_{S^4}(g,\theta)= \int d\mu_\alpha \,e^{-\frac{S_{cl}(\alpha)}{g^2}} \vert Z_{1-loop}(\alpha)\vert^2\, \vert Z_{inst}(\alpha,\tau)\vert^2\,,
\end{eqnarray}
where $g$ is the gauge coupling and $\theta$ the topological angle, while $d \mu_\alpha$ is the measure over the Cartan subalgebra of the gauge group and $Z_{inst}(\alpha,\tau)$ denotes Nekrasov \cite{Nekrasov:2002qd,Nekrasov:2003rj} instanton partition function with $\tau = \frac{i}{g^2} +\frac{\theta}{2\pi}$.

The vortex partition function is now replaced by Nekrasov partition function, the $3$-d FI parameter translates into the $4$-d coupling $1/g^2$, and the discrete sum of (\ref{factorisation}) becomes an integral over the Cartan subalgebra of the gauge group.

As a concrete example let us consider pure $\mathcal{N}=2$ with gauge group $SU(2)$ so that the integral over the Cartan subalgebra reduces to an integral over $\alpha\in\mathbb{R}$. In this case following \cite{Aniceto:2014hoa} we can rewrite the path-integral in the topological sector form 
\begin{equation}
Z^{SU(2)}_{S^4}(g,\theta)=  \sum_{B\in\mathbb{Z}} e^{-\frac{2\pi}{g^2}  | B|+ \frac{i \theta}{2\pi} B} 
\zeta_B(g)\,,
\end{equation}
where for $B\geq0$ we have
\begin{equation}
\zeta_B(g) = \sum_{N\geq 0 } e^{-\frac{4\pi}{g^2} N} \int_{-\infty}^{\infty} d\alpha \,e^{-\frac{S_{cl}(\alpha)}{g^2}} \vert Z_{1-loop}(\alpha)\vert^2  \, Z_{inst}^{(B+N)}(i\alpha) \,Z_{inst}^{(N)}(-i \alpha)\,,
\end{equation}
while for $B<0$ we just need to take the complex conjugate of this. 
These two equations should be compared with their $3$-d counterparts (\ref{eq:TopoTS}) and (\ref{eq:TSN}).
Note the function $Z_{inst}^{(k)}(i\alpha)$ corresponds to the $k$-instanton Nekrasov partition function, e.g. $Z_{inst}^{(0)}(i\alpha)=1$, and can be explicitly found in \cite{Aniceto:2014hoa} for $k\leq 8$.

As for the three dimensional case, in this $S^4$ example we have some extra structure.
It is clear from the argument outlined above that if we were able to compute with resurgence methods from the purely perturbative expansion, i.e. $B=0,N=0$ above, all the contributions from the instanton-anti-instanton sectors, i.e. $B=0,N>0$, we would then be able to calculate all the perturbative and non-perturbative data in all the other topological sectors.

The resurgence analysis for this class of $\mathcal{N}=2$ with different matter content has been discussed in details in \cite{Aniceto:2014hoa} (see also the earlier \cite{Russo:2012kj}) and the authors showed that it is not however possible to reconstruct in this way the instanton-anti-instanton sectors from perturbation theory. The singularities of the Borel transform for the purely perturbative sector are not directly related to instanton-anti-instanton configurations. It was subsequently realised \cite{Honda:2017qdb} (at least for the three dimensional case) that these singularities come from new finite action complexified supersymmetric solutions.

The reason for this is subtle: although the Borel transform of the perturbative series has poles, these are coming from the one-loop determinant of matter multiplets, i.e. hypermultiplets of $\mathcal{N}=2$, and the fields involved in these complexified supersymmetric solutions come precisely from the matter sector. However we know that instantons are present even in absence of hypermultiples. This suggests that the instanton-anti-instanton poles are hidden by a Cheshire Cat structure inside the one-loop determinant of the $\mathcal{N}=2$ vector multiplet. We have analysed the localised one-loop determinant for the vector multiplet but failed so far to find a suitable Cheshire Cat deformation that would allow us to carry on the programme outlined above.

It would be extremely interesting to see if these kind of holomorphic/anti-holomorphic structures, intertwining the complexified coupling constant with the theta angle, can be extended to less supersymmetric theories as for example just pure Yang-Mills, thus allowing us to extend resurgence to the whole triangle of Figure \ref{DunneUnsal}. 

\section{Conclusion}\label{conclusion}
 
In this paper we have considered the partition function for abelian $\mathcal{N}=2$ supersymmetric theories with different matter content living on a squashed $S^3$. This problem was first analysed in \cite{Fujimori:2018nvz} for $\mathcal{N}=2$ Chern-Simons matter theories where the authors showed that the presence of Stokes phenomenon in the thimble decomposition was directly related to the ambiguities in resummation of the asymptotic perturbative expansion in the small coupling $g=1/k$, with $k$ the Chern-Simons level.

In our work we have set to zero the Chern-Simons level and considered the perturbative expansion in large FI parameter. 
Firstly we have analysed the Picard-Lefschetz decomposition of the localised path-integral into steepest descent contours and we have shown that if a suitable complexification of the squashing parameter $b$ is introduced, a hidden topological angle seems to appear and a steepest descent contour can be associated to each topological sector.

Physically this complexified squashing parameter can be seen as adding a chemical potential for rotation of the $S^2$ so that vortices and anti-vortices rotate oppositely. The FI term on the other hand regulates the size of the vortices localised at north and south poles, thus we have a play off between these two parameters.

As we vary the complexified squashing $b$ and the Fayet-Iliopoulos we observe the splitting of saddle points into different topological sectors as well as Stokes phenomenon whenever a saddle crosses the steepest descent cycle coming from another saddle. For large enough FI parameter these saddles can be associated to point-like vortex (or anti-vortex) solutions and the path-integral can be decomposed into a sum of contour integrals, one for each topological sector. 

Having split the path integral into a sum over topological sector we first perform a semi-classical expansion $\xi \gg 1$ showing that, due to the supersymmetric nature of the observable under consideration, every perturbative series truncates after finitely many orders. 
These $\mathcal{N}=2$ theories provide another interesting example of a field theory that lies at a very special point in theory space where lots of miraculous cancellations hide the resurgence structure rendering the perturbative expansions in each of the non-perturbative sectors as truncating series. 

To use the resurgence machinery we then introduce a Cheshire Cat deformation by analytically continuing the number of chiral fields. As soon as the deformation parameter is generic we immediately re-introduce the asymptotic nature of perturbation theory. Thus we work at a generic point and using resurgent analysis reconstruct the vortex-anti-vortex contributions from the deformed, factorially growing and purely perturbative data. Once the deformation parameter is set back to its physical value we have that the asymptotic tail of the perturbative series vanishes but the non-perturbative contributions still stand.

We also comment on the strange nature of the other topological sectors which should be in principle completely disconnected from perturbation theory and the trivial topological sector but in practice they are not.
This suggests the existence of additional structures, beyond the standard resurgence framework, namely what is called vortex/anti-vortex factorisation. Similar structures are present also in $2$-d and $4$-d supersymmetric theories and allow us to use non-perturbative data in the topologically trivial sector to obtain non-perturbative data in other topological sectors. 

In particular we pose the question on how to extend this Cheshire Cat deformation combined with Nekrasov partition function to the case of say the pure $SU(2)\, \mathcal{N}=2$ supersymmetric theory on $S^4$ where on the one hand we do expect infinitely many instanton-anti-instanton contributions but on the other hand these are somehow completely hidden from perturbation theory.

\acknowledgments

The authors would like to thank Teresa Bautista, Mathew Bullimore, Stefano Cremonesi, Gerald Dunne, Masazumi Honda, Axel Kleinschmidt, Tatsuhiro Misumi, Matthew Renwick and Mithat \"Unsal for useful discussions. D.D. thanks the Albert Einstein Institute and in particular Hermann Nicolai for the hospitality and support during the final stages of this project.

\appendix
\section{Double Sine Function Identities}\label{doublesine}

We here state a number of useful formulae for the double sine function. The Appendices of \cite{Pasquetti:2011fj,Benini:2011mf} contain comprehensive lists of properties for this function; otherwise we refer to \cite{Rujisenaars,Kurokawa,DeBult}.

We define the double sine function as
\begin{eqnarray}\label{sbdefinition1}
s_b\left(x\right) &=& \prod_{m,n\geq0}\frac{(mb+n/b+Q/2-ix)}{(mb+n/b+Q/2+ix)}\;\;\; .
\end{eqnarray}

Following \cite{DeBult} we can introduce the double gamma function $\Gamma_2$ defined by the analytic continuation
\begin{equation}
\Gamma_2(z;\omega_1,\omega_2) =\exp\left[ \partial_s\left.\left( \sum\limits_{m,n\geq0} (m \omega_1+n\omega_2+z)^{-s}\right)\right\vert_{s=0}\right]\,,
\end{equation}
so that the formal infinite product (\ref{sbdefinition1}) can be rewritten as
\begin{equation*}
s_b(x+i Q/2) = \frac{\Gamma_2(ix; b,b^{-1})}{\Gamma_2(Q-ix;b,b^{-1})} =\Gamma_h(ix;b,b^{-1}) = S_2(i x\vert b,b^{-1})^{-1} = G(-i b,-ib^{-1};ix - Q/2)\;,
\end{equation*}
where $\Gamma_h$ is van de Bult \cite{DeBult} hyperbolic gamma function, $S_2$ is the double Sine function of Kurokawa and Koyama \cite{Kurokawa} and $G$ is Ruijsenaars hyperbolic gamma \cite{Rujisenaars}.

Obviously $s_b(0)=1$ and furthermore $s_b(x) $ has zeroes on the lattice $\Lambda_+ =-iQ/2 -i b \,\mathbb{Z}_{\geq 0} -i/b\, \mathbb{Z}_{\geq 0}$ and poles on the lattice  $\Lambda_- = +i Q/2 +i b\, \mathbb{Z}_{\geq 0} +i/b\, \mathbb{Z}_{\geq 0}$. Both the zeroes and poles are simple provided that $b^2$ is not rational.
In particular the pole at $x = iQ/2$ is always simple and we have
\begin{equation}
\label{eq:Residue}
s_b(x) = \frac{i}{2\pi (x-i Q/2)}+O(1)\,,\qquad\qquad x\to iQ/2\,.
\end{equation}
From the known functional equations
\begin{align}\label{eq:functional}
s_b(x+ib/2) s_b(-x+ib/2) &= \frac{1}{2\cosh(\pi b x)}\,,\\
s_b(x+iQ/2) &= \frac{s_b(x+iQ/2-ib)}{2i \sinh(\pi b x)}\,,
\end{align}
one can derive the general expressions
\begin{align}
\frac{s_b(x+ i Q/2 + i m b+ i n /b)}{s_b(x+ i Q/2)} &\label{eq:functional2}= \frac{(-1)^{mn}}{\prod\limits_{k=1}^{n}2i\sinh[\pi b(x+i k b)] \prod\limits_{l=1}^{m}2i\sinh[\pi/b(x+i l/b)]}\,,\\
\frac{s_b(x-i Q/2 + i m b+ i n /b)}{s_b(x- i Q/2)} &\notag= \frac{(-1)^{mn}}{\prod\limits_{k=1}^{n}2i\sinh[\pi b(x-i Q+i k b)] \prod\limits_{l=1}^{m}2i\sinh[\pi/b(x-iQ+i l/b)]}\,,
\end{align}
allowing us to obtain the residue at different poles from the residue at zero (\ref{eq:Residue}).

A useful infinite product identity for $s_b(x)$ is given by
\begin{eqnarray}\label{SaraRatio}
s_b(x)&=&e^{-i\pi\frac{x^2}{2}- i \pi\frac{b^2+b^{-2}}{24}}\frac{\prod\limits_{k=0}^\infty\left(1+e^{2\pi bx}e^{2\pi i b^2(k+1/2)}\right)}{\prod\limits_{k=0}^\infty\left(1+e^{2\pi x/b}e^{-2\pi i(k+1/2)/b^2}\right)} \;\;\;  ,
\end{eqnarray}
which can be regularised using q-Pochhammers symbols. Recall that the q-Pochhammer $(a;q)_\infty$ is defined as
\begin{eqnarray}
(a;q)_\infty = \prod\limits_{k=0}^\infty\left(1-a q^k\right) \;\;\; .
\end{eqnarray}
Using this we can thus write
\begin{eqnarray}\label{qpochsb}
s_b(x)=e^{-i\pi\frac{x^2}{2}- i \pi\frac{b^2+b^{-2}}{24}}\frac{\left(-e^{2\pi bx+\pi ib^2};e^{2\pi ib^2}\right)_\infty}{\left(-e^{2\pi x/b-\pi i/b^2};e^{-2\pi i/b^2}\right)_\infty} \;\;\; ,
\end{eqnarray}
valid for $\mbox{Im}( b^2 )>0$ so that $ \vert e^{2\pi ib^2}\vert <1$ as well as $\vert e^{-2\pi i/b^2}\vert <1$ .

\bibliography{cheshire.bib}
\end{document}